\documentclass[sigconf]{acmart}
\AtBeginDocument{%
  }
\usepackage{subfigure}  
\usepackage{multirow}
\usepackage{tabularx}
\usepackage{booktabs}
\usepackage{longtable}
\usepackage{array}
\usepackage{subcaption}
\usepackage{graphicx}
\usepackage{circledsteps}
\usepackage{hyperref}
\usepackage{soul}

\hypersetup{
    pdftitle={VisceroHaptics: Investigating the Effects of Gut-based Audio-Haptic Feedback on Gastric Feelings and Gastric Interoceptive Behavior},
    pdfauthor={Anonymous Author(s)},
    pdfsubject={Your Document Subject},
    pdfkeywords={embodied interaction, input techniques}
}

\setcopyright{acmlicensed}
\copyrightyear{2026}
\acmYear{2026}
\setcopyright{cc}
\setcctype{by-nc-nd}
\acmConference[CHI '26]{Proceedings of the 2026 CHI Conference on Human Factors in Computing Systems}{April 13--17, 2026}{Barcelona, Spain}
\acmBooktitle{Proceedings of the 2026 CHI Conference on Human Factors in Computing Systems (CHI '26), April 13--17, 2026, Barcelona, Spain}
\acmPrice{}
\acmDOI{10.1145/3772318.3790268}
\acmISBN{979-8-4007-2278-3/2026/04}

\newcommand{\workingcolor}{black}

\newcommand{\change}[1]{\textcolor{\workingcolor}{#1}}
\newcommand{\revise}[1]{\textcolor{black}{#1}}

\begin{document}


\title[VisceroHaptics]{VisceroHaptics: Investigating the Effects of Gut-based Audio-Haptic Feedback on Gastric Feelings and Gastric Interoceptive Behavior}

\author{Mia Huong Nguyen}
\orcid{1234-5678-9012}

\affiliation{%
  \department{Augmented Human Lab, School of Computing}
  \institution{National University of Singapore}
  \city{Singapore}
  \country{Singapore}
}
\email{mia@ahlab.org}

\author{Jochen Huber}
\authornote{Both authors contributed equally to this work}
\affiliation{%
  \department{Faculty of Industrial Technologies}
  \institution{Furtwangen University}
  \city{Furtwangen}
  \country{Germany}
}
\email{jochen.huber@hs-furtwangen.de}

\author{Morith Alexandra Messerschmit}
\authornotemark[1]
\affiliation{%
\department{Augmented Human Lab, School of Computing}
  \institution{National University of Singapore}
  \city{Singapore}
  \country{Singapore}
}
\email{moritz@ahlab.org}

\author{Suranga Chandima Nanayakkara}
\affiliation{%
\department{Augmented Human Lab, School of Computing}
  \institution{National University of Singapore}
  \city{Singapore}
  \country{Singapore}
}
\email{suranga@ahlab.org}

\renewcommand{\shortauthors}{Nguyen et al.}
\newcommand{\todo}[1]{\textcolor{red}{#1}}
\newcommand{\move}[1]{\textcolor{black}{#1}}

\ccsdesc[300]{Human-centered computing~Interaction paradigms}
\ccsdesc[300]{Human-centered computing~Interaction devices}

\keywords{Gut-brain connection, gastric interoception, gut feelings}
\begin{teaserfigure}
  \includegraphics[width=\textwidth]{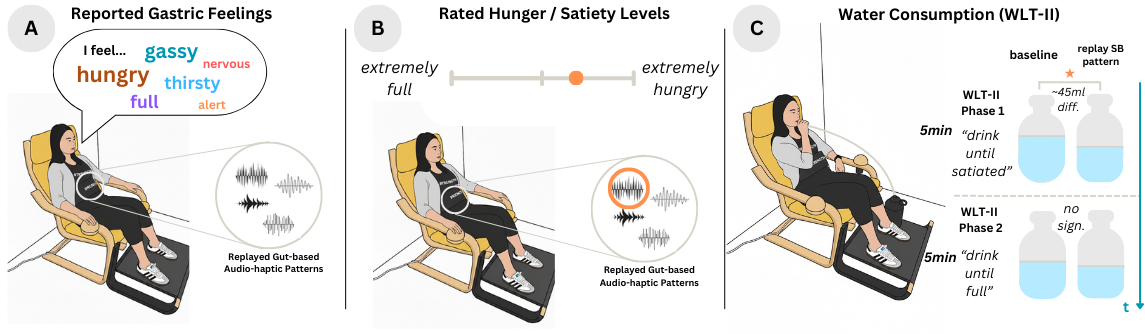}
  \caption{We report on three empirical studies examining how gut-based audio-haptic feedback affects gastric feelings and interoceptive behavior. Using bowel sound recordings from participants performing a two-phase water load test (WLT-2), we conducted: (A) a study on the effects of gut-based audio-haptic feedback on users' feelings; (B) an investigation of how such patterns influence users' perception of hunger and satiety levels; and (C) an assessment of their effects on drinking behavior. Results show that single-burst patterns significantly increased water consumption in a within-subject design, suggesting that non-invasive gut-based audio-haptic feedback can influence gastric interoceptive behavior.}
  \Description{Three-panel diagram (A, B, C) illustrating experimental setup. Panel A shows reported gastric feelings with audio patterns, Panel B shows hunger/satiety rating scale, Panel C shows water consumption protocol with WLT-II phases}
  \label{fig:teaser}    
\end{teaserfigure}

\begin{abstract}
Gastric interoception influences eating behavior and emotions, making its modulation valuable for healthcare and human-computer-interaction applications. However, whether gastric interoception can be modulated noninvasively in humans remains unclear. While previous research indicates that abdominal-sound-driven haptic feedback resembles gut sensations, its impact on gastric feelings and gastric interoceptive behavior is unknown. We conducted three experiments totalling 55 participants to investigate how gut-sound-driven audio-haptic feedback applied to the stomach (1) affects user's feelings (2) influences perception of hunger and satiety levels and (3) influences gastric interoceptive behavior, quantified with Water Load Test-II. Results revealed that audio-haptic feedback patterns (a) induced the feelings of hunger, fullness, thirst, stomach upset, (b) increased hunger level, and (c) significantly increased volumes of ingested water. This work provides the first evidence that audio-haptic stimulation can alter gastric interoceptive behavior, motivating the use of noninvasive methods to influence users' feelings and behaviors in future applications.
\end{abstract}
\maketitle

\section{Introduction}

\textit{"The way to one's heart is through the stomach" } --- ancient proverb. While this wisdom may not be anatomically precise, recent scientific advances suggest there is some truth to it~\cite{morais_Gut_2021, omahony_Serotonin_2015, margolis_microbiota-gut-brain_2021}. The gut-brain axis is now widely accepted as a fundamental connection, influencing various aspects of a person's life, from eating behaviors to emotional experiences \cite{mayer_gut_2011, appleton_gut-brain_2018, holzer_interoception_2017}. 
Signals from the gut influence feelings of hunger and fullness, which in turn shape our appetite and eating habits \cite{ nakazato_role_2001, amin_hunger_2016,berthoud_vagal_2008,heisler_appetite_2017}. These gut signals are sent to the brain through multiple pathways, including both hormonal and neurological channels \cite{kaelberer_gut-brain_2018}. The collective signals sent from the gut to the brain through these  pathways are referred to as gastric interoception \cite{mayer_gut_2011, holzer_interoception_2017}.

Modulating gastric interoception therefore holds large potential for treating health-related conditions \cite{srinivasan_vibrating_2023, ramadi_bioinspired_2023, payne_bioelectric_2019, mccallum_mechanisms_2010, tiemann_relevance_2025}. It also opens up new avenues for research, deepening our understanding of the gut-brain connection \cite{noauthor_stomachs_2025}.
From a Human-Computer Interaction (HCI) perspective, modulating gastric interoception introduces a new experiential dimension in how users interact with technology \cite{nguyen_gutio_2025} and \change{enables HCI researchers to explore novel gut-based interfaces that communicate with users through their visceral sensations, creating intimate and embodied experiences that extended beyond traditional modalities}. By leveraging the gut's inherent connection to emotional and decision-making processes, such interfaces could potentially influence user behavior in subtle, yet meaningful ways---from promoting healthier eating patterns to enhancing emotional regulation in digital wellness applications. 

Existing work in the field of biotechnology \cite{srinivasan_vibrating_2023, ramadi_bioinspired_2023} attempted to modulate gastric signals through invasive devices such as ingestible pills \cite{srinivasan_vibrating_2023, ramadi_bioinspired_2023} and gastric-electrical stimulation (GES) \cite{payne_bioelectric_2019}. Ingestible pills directly stimulate the cells inside the stomach's mucosa to promote or inhibit the production of ghrelin, the "hunger hormone," and modulate the feeling of hunger \cite{ramadi_bioinspired_2023} and fullness \cite{srinivasan_vibrating_2023} respectively. These pills represent a direct approach to modulate gastric interoception by modifying gut signals in the hormonal pathway. However, the effectiveness of these pills was only tested in pigs. Gastric-electrical stimulations can also be leveraged to modulate gastric interoception by modifying the signals in the neurological pathways. For instance, such stimulations have been used to treat patients with recalcitrant gastroparesis -- \change{severe form of the stomach-emptying disorder}, but requires the device to be implanted inside patients' bodies \cite{payne_bioelectric_2019}. Though proven to be effective in modulating gastric interoceptions, these methods are invasive, and their use is limited in medical settings. 
%

Direct modification of gastric signals through hormonal or neurological pathways, while theoretically sound, remains impractical for interactive technologies due to invasive implementation requirements. Instead of direct modification of in-body gastric signals, we augment gastric sensations to modulate hunger perception by replaying gut-based feedback cutaneously through an external device. This way, gastric interoception can be modulated non-invasively. Our line of work is inspired by theories of grounded cognition, which emphasize that bodily sensations play a crucial role in shaping cognition \cite{barsalou_grounded_2008}. Studies showed that false heart-rate feedback altering perceived heart rate \cite{costa_emotioncheck_2016} influenced anxiety \cite{costa_boostmeup_2019, costa_emotioncheck_2016}, arousal \cite{valins_cognitive_1966} and task performance\cite{costa_boostmeup_2019, valins_cognitive_1966}. These findings suggest that convincing manipulation of bodily feedback creates new avenues for behavioral influence. Extending this framework to gastric perception, we hypothesize that modulating perceived gastric activity offers similar potential for behavioral modification. Early work by \citet{nguyen_gutio_2025} showed that presenting gut sounds as haptic feedback on the abdomen is perceived as realistic and resembles gut sensations that users \change{had} experienced in the past. It remains unclear: (1) how gut-based \change{audio-haptic} feedback influences users’ feelings in the here and now (RQ1), (2) how it shapes perceptions of both hunger and satiety level (RQ2), and (3) whether gastric interoceptive behavior can be modulated through gut-based audio-haptic feedback (RQ3).

To address these questions sequentially, we conducted a series of three studies with a total of 55 participants. 
%
Addressing RQ1, we empirically investigated how 18 users experience audio-haptic stimuli, reporting feelings every minute. Results showed gut-based audio-haptic feedback induced hunger, fullness, and stomach upset, as well as emotions like anxiety and anticipation. \change{Varying} strengths of hunger and fullness feeling (e.g. slightly hungry vs. very hungry) may significantly influence gastric interoceptive behavior. \change{To this end,} study 2 quantifies the impact of audio-haptic \change{stimuli} and addresses RQ2 by asking 16 participants to rate their hunger and fullness levels after experiencing \change{four stimuli}. Our analysis identified \change{two} stimuli that significantly differed in influencing perceptions of hunger and satiety levels. Lastly, Study 3 investigates RQ3 by examining whether the perceived hunger changes translate \change{into} gastric interoceptive behavior changes. \change{The latter was} quantified by the amount of water participants drink during a two-stage Water Load Test (WLT-II) \cite{van_dyck_water_2016}. In this within-subject, 3-session study, 21 participants performed the WLT-II test under gut-based audio-haptic feedback conditions. We found one haptic condition significantly \change{influenced gastric interoceptive behavior}.

To the best of our knowledge, our results provide the first empirical evidence showing that non-invasive gut-based audio-haptic stimulation can influence gastric interoceptive behavior. The audio-haptic stimulus that participants least attributed to as hunger-inducing in our first and second studies turned out to be the hunger-inducing stimulus in our third study. In the latter, we saw a significantly higher water intake compared to the baseline condition, where no stimulus is activated. This suggests a dissociation between subjective hunger perception and gastric interoceptive behavioral responses. Our work thus makes the following contributions:
\begin{itemize}
    \item Empirical findings on the effect of gut-based audio-haptic feedback on users' feelings.
    \item Empirical findings on the effect of gut-based audio-haptic feedback on users' perceived hunger and satiety levels.
    \item Empirical evidence that shows the possibility of altering gastric interoceptive behavior through non-invasive audio-haptic feedback.
\end{itemize}


\section{Related Work}
\change{Our work contributes to the growing body of knowledge on interoception manipulation and the working mechanisms of bowel sounds. We review prior work within those fields in the following sections.}
\subsection{\change{Sensory Feedback for Altering Body and Affective Perception}}
\change{
Since the discovery of the rubber hand illusion \cite{botvinick_rubber_1998}, researchers have increasingly investigated the effect of different kinds of stimuli on human perception. In the HCI literature, prior works have examined the effect of visual, auditory, and haptic feedback on users' perception of their body \cite{tajadura-jimenez_as_2019, preatoni_reshaping_2023, ley-flores_effects_2022, stanton_feeling_2017, tajadura-jimenez_as_2015, moullec_multi-sensory_2022}. For example, \citet{ tajadura-jimenez_as_2019} found that lighter footstep sounds make people feel as if their body was lighter; \citet{piryankova_can_2014} found that the shape and cloth texture of virtual reality avatars can affect users' perception of their Body Mass Index (BMI); \change{\citet{jain_Frisson_2022} induced aesthetic chills using haptic feedback, while \citet{moullec_multi-sensory_2022} found that combining visual and haptic cues can increase the sensation of effort in VR environment. Collectively, this prior art suggests that our bodily self-perception is malleable and can be augmented through multi-sensory feedback. Our work expands this literature by introducing a new technique to augment gastric feelings.} }

\subsection{\change{Interoception Manipulation via Gut}}
\change{Existing studies in the literature suggest the potential of interoceptive manipulation for health and well-being. However, the majority of the explorations are limited to cardiac \cite{costa_emotioncheck_2016, costa_boostmeup_2019, valente_modulating_2024, hassan_Heartbeat_2025} and respiratory signals \cite{ghandeharioun_brightbeat_2017, moraveji_peripheral_2011, miri_evaluating_2020, brooks_augmented_2024}. While valuable, these interoceptive signals cannot replace gastric feelings \cite{critchley_interoception_2017}. For example, the feelings of a growling stomach during hunger or the bubbling sensation due to bloatedness cannot be fully replicated or induced through modulating cardiac or respiratory signals alone.}

\change{
While there has been a growing interest in the gastrointestinal (GI) tract as a novel site of interaction \cite{li_Guts_2018, vujic_going_2020, pasumarthy_gooey_2021, pasumarthy_gooey_2022, pasumarthy_go-go_2024},  prior work has focused on aspects other than manipulating gastric interoception, such as developing wearable sensing devices \cite{nguyen_catch_2024, vujic_gut-brain_2019}, building knowledge and awareness \cite{pasumarthy_gooey_2021, pasumarthy_gooey_2022, pasumarthy_go-go_2024}, or building novel experience with ingestible pill~\cite{li_Guts_2018}. The space of gastric interoception manipulation remains largely unexplored within the HCI community. The only prior work on gastric interoception is by \citet{nguyen_gutio_2025}, who investigated gut-based haptic feedback\footnote{\change{Note that this feedback is haptic-only, where as our feedback is audio-haptic}}to induce gastric sensations in users.
They found that when applied to users' stomachs, abdominal sound patterns were perceived as realistic and resembled past gastric experiences, as if originating from within the body.
While promising, this work was limited to retrospective perception of gastric sensations—how the induced sensations resembled past experiences rather than their immediate effects on current feelings. Furthermore, the translation of perceived gastric changes into behavioral modifications was not investigated. Our work contributes to the interoception literature by demonstrating how gut-based audio-haptic feedback can be designed to augment gastric feelings and sublimally influences behaviors.
}
\begin{figure*}[t]
    \centering
    \includegraphics[width=1\linewidth]{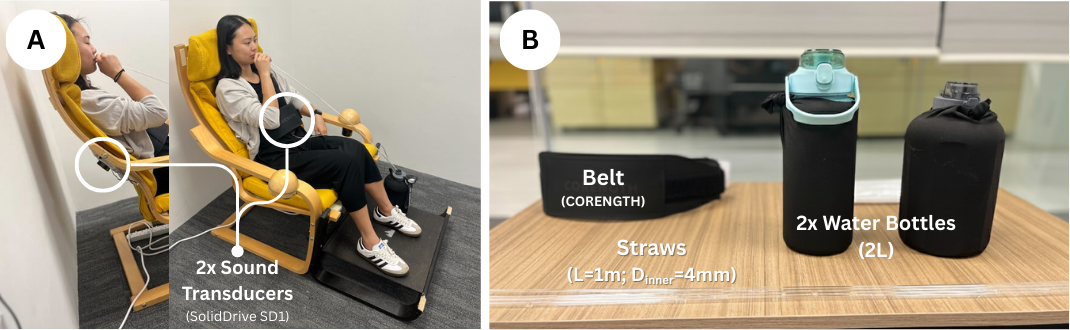}
    \caption{(A) Study setup with a participant drinking water through a straw while gut sounds are being played back to the participant via two sound transducers (SolidDrive SD1) attached to the back of a chair and the participant's belly. (B) Apparatus components consisting of a CORENGTH Belt to strap the sound transducer to the participant's belly, two 2L water bottles, and several straws with a length of 1m and an inner diameter of 4mm.}
    \label{fig:setup_and_apparatus}
    \Description{Two-panel photograph showing experimental equipment. Panel A shows participants in reclined chairs with sound transducers, Panel B shows equipment including a belt, straws, and 2L water bottles on a table.}
\end{figure*}
\subsection{\change{Bowel Sound Patterns \& Characteristics}}
\change{ The movements of the intestines create bowel sounds \cite{ferguson_inspection_1990}. Prior literature documented four types of bowel sounds, namely, single burst (SB), multiple burst (MB), continuous random sound (CRS) and long harmonic sound (LHS) \cite{nguyen_gutio_2025,du_mathematical_2018, li_bowel_2012, dimoulas_Pattern_2011}. SB sounds are characterized by solitary pulses that may have tonal sounding or dry explosive nature \cite{dimoulas_Pattern_2011}, MB sounds can be described as a sequence of solitary pulses with an interval time between each pulses smaller than 100 ms \cite{du_mathematical_2018}. CRS sounds are continuous without any significant silent gaps and have little rhythms and finally, LHS sounds are whistling-like sounds and have one to a dozen frequency components in the
spectrogram \cite{du_mathematical_2018, dimoulas_Pattern_2011}.} 

\change{Bowel sounds' characteristics are often described in terms of duration, and acoustic features such as spectral centroid and spectral flatness \cite{du_mathematical_2018}. 
Duration describes how long a single sound instance lasts, spectral centroid refers to the fundamental frequency of the sound instance, and spectral flatness describes the degree to which the sound is tonal versus noise-like (with high flatness indicating noise-like characteristics and low flatness indicating tonal characteristics with prominent harmonic content)~\cite{du_mathematical_2018, herre_Robust_2002}.
These bowel sound characteristics typically do not change across time \cite{ranta_Digestive_2010}. However, their frequency (i.e., how often they occur in a given time unit) may vary due to the effect of food intake \cite{horiyama_bowel_2021}.} 
\change{In addition, as these bowel sounds propagate through the GI tract, they generate vibrations perceivable by the mechanoreceptors lining the inner layer of the GI tract \cite{furness_Gut_2013,mercado-perez_gut_2022} and by those on the skin surface. Thus, humans can consciously perceive bowel sounds not only through the auditory channel but also as tactile sensations deep inside the stomach.}

\subsection{Quantifying Gastric Interoceptive Behavior}
\label{section:wlt}
To objectively quantify gastric interoception, the Water Load Test-II (WLT-II) by \citet{van_dyck_water_2016} is widely used \cite{salaris_investigating_2025, tiemann_relevance_2025}. Participants are required to fast for 3 hours and refrain from drinking for 2 hours before the test. They are then asked to drink non-carbonated water at room temperature \change{ in 2 successive 5-min phases. The protocol runs as follows:} In the first phase, participants are instructed to drink water until they perceive signs of satiation, that is the sensation that determines meal termination. For this phase, participants are given the following instructions: \textit{"During the following five minutes, we ask you to drink water until perceiving signs of satiation. By satiation we mean the comfortable sensation you perceive when you have eaten a meal and you have eaten enough, but not too much."} \change{The volume of consumed water until the participant felt satiated is recorded at the end of phase 1 ($sat\_ml$).} In the second phase, participants are instructed to drink water until reaching the point of maximum fullness. In this case, participants read the following instructions: \textit{"We now ask you to drink again until your stomach is completely full, that is, entirely filled with water. You have five minutes to do this"}. \change{The total volume of consumed water is recorded at the end of phase 2 ($total\_ml$).}  Participants are asked to drink water through long straws from non-transparent flasks to control their drink speed, as well as to blind them from knowing how much water they have drunk. The ratio of water volume drunk until satiation over the total volume drunk is calculated as $sat\_\% = sat\_ml/total\_ml$.  \change{
In our work, we utilize WLT-II to measure the impact of gut-based audio-haptic stimuli on gastric interoception and answer RQ3.}

\section{Study 1: Effect of Gut-based Audio-haptic Feedback on Gastric Feelings}

Gut sounds are audible sounds produced by the GI tract, experienced both as internal tactile sensations and external auditory feedback. Building on research in gut-based haptic feedback and body perception manipulation with auditory feedback, this study addresses \textbf{RQ1:} \textit{\change{"How does gut-based audio-haptic feedback influences users' feelings in the here and now?"}}

\change{Since gut sounds are experienced through both auditory and tactile channels, }we preserve both modalities to create comprehensive gut-based audio-haptic feedback. We recorded bowel sounds from participants during the WLT-II task (see section~\ref{section:wlt}) and applied established bowel sound analysis methods\cite{du_mathematical_2018, vasseur_postprandial_1975, allwood_advances_2019} to develop our audio-haptic stimuli.
The following sections describe our stimulus curation process, study procedure, results, and insights that informed Study 2's design.  \change{All studies are approved by the university's Human Ethics Committee (NUS-IRB-2024-586). }

\begin{figure*}
    \centering
    \includegraphics[width=1\linewidth]{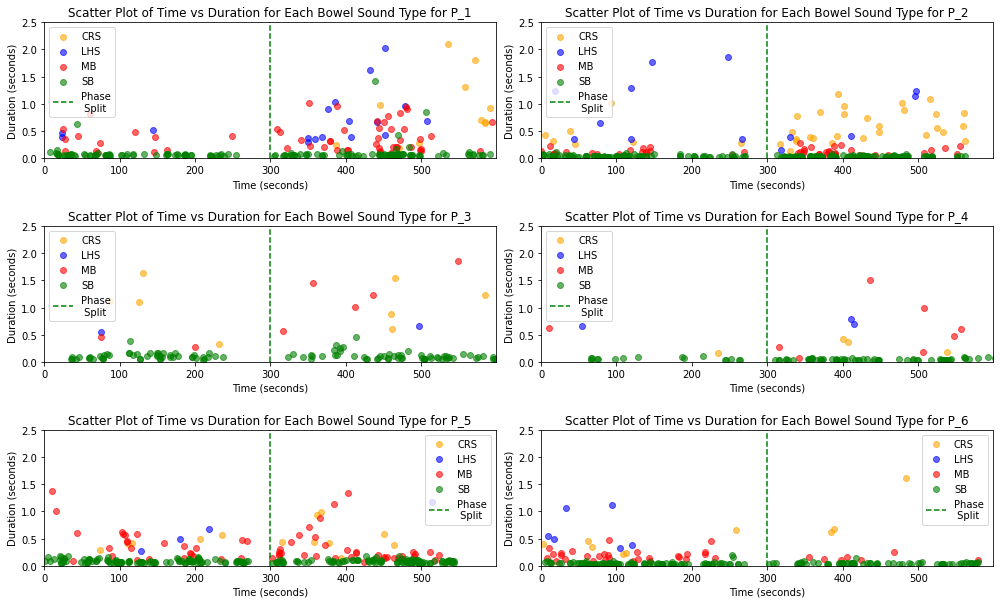}
    \caption{Scattered plots of bowel sounds distribution over time and their duration. CRS: Continuous Random Sounds, LHS: Long Harmonic Sounds, MB: Multiple Bursts, SB: Single Bursts}
    \label{fig:bowel_sound_scattered}
    \Description{Six scatter plots (P_1 through P_6) showing time vs duration for different bowel sound types (crs, lhs, mb, sb), with color-coded data points representing each sound type.}
\end{figure*}
\subsection{Stimuli Creation}

Early investigation by \citet{nguyen_gutio_2025} showed that bowel sound patterns, when transformed into haptic feedback, were perceived as realistic, coming from the inside, and resembled gastric feelings. Building on this knowledge, we created a pool of stimulus spanning these 4 categories by performing the following steps: 

\begin{table*}[hbt!]
\centering
\resizebox{\textwidth}{!}{
\color{\workingcolor}
\begin{tabular}{l|cccc|ccc|ccc|ccc}
\hline
& \multicolumn{4}{c|}{\textbf{Count}} & \multicolumn{3}{c|}{\textbf{Duration (seconds)}} & \multicolumn{3}{c|}{\textbf{Mean Spectral Centroid (Hz)}} & \multicolumn{3}{c}{\textbf{Mean Spectral Flatness}} \\
\cline{2-14}
& Phase 1 & Phase 2 & Total & \% & Phase 1 & Phase 2 & Temporal  & Phase 1 & Phase 2 & Temporal  & Phase 1 & Phase 2 & Temporal \\
\textbf{Label} & & & & & Mean (Std) & Mean (Std) &corr. & Mean (Std) & Mean (Std) &corr. & Mean (Std) & Mean (Std) &  corr. \\
\hline
CRS & 27 & 60 & 87& 6.48 & 0.58 (0.54) & 0.63 (0.34) & -0.001 & 538.19 (130.08) & 610.77 (143.69) & -0.122 & 7.2E-05 (8.0E-05) & 6.6E-05 (7.5E-05) & 0.021 \\
LHS & 36 & 31 & 67 & 4.99& 0.85(0.59) & 0.72 (0.44) & -0.011 & 520.75 (106.70) & 452.10 (205.54) & 0.010 & 9.3E-05 (1.1E-04) & 9.1E-05 (1.8E-04) & -0.024 \\
MB & 90 & 129 & 219 & 16.31 & 0.33 (0.33) & 0.34 (0.34) & 0.075 & 723.52 (267.65) & 712.22 (226.72) & 0.197 & 1.5E-04 (2.5E-04) & 1.4E-04 (1.9E-04) & -0.025 \\
SB & 449 & 521 & 970 & 72.23 & 0.06 (0.05) & 0.06 (0.04) & -0.002 & 922.69 (338.31) & 952.33 (336.15) & 0.031 & 4.6E-04 (8.7E-04) & 4.8E-04 (6.6E-04) & 0.002 \\
\hline
\end{tabular}
}
\caption[Acoustic features comparison across phases by bowel sound types]{\change{Comparison of acoustic features across phases by bowel sound type. Sound's characteristics are comparable between the two phases while the frequency of occurrence (count) differs. Temporal correlations are calculated by computing the Pearson correlation between the value of interest and the sounds occurrences' timestamps}.}
\label{tab:descriptive_stats}
\end{table*}

\textbf{Step 1: Bowel sound recording}. We collected bowel sounds from 7 participants performing the WLT-II through a pilot study. Participants fasted for 3 hours and abstained from water for 2 hours before arrival.
Upon arrival, researchers positioned a recording device containing an embedded electrical stethoscope on participants' lower left abdominal quadrant using a sports belt. The stethoscope position and belt tension were calibrated to optimize recording quality. After explaining the WLT-II procedure (detailed in Section \ref{section:wlt}), participants were asked to remain silent and minimize movement to prevent audio contamination. Bowel sounds were recorded throughout the 10-minute drinking period. 

\textbf{Step 2: Bowel sound segmentation and analysis.}
After collecting bowel sounds from 7 participants, we verified the presence of all 4 bowel sound categories in our dataset. Following the categorization framework from Du et al. (2018)\cite{du_mathematical_2018}, the first author manually identified and labeled bowel sounds across all 7 recordings. This analysis yielded 970 SB, 219 MB, 88 CRS, and 66 LHS. Examples of bowel sounds in these categories are included in Figure \ref{fig:example}, Appendix A. We analyzed the acoustic characteristics of each category and examined correlations between sound properties and temporal occurrence. These findings are summarized in Table \ref{tab:descriptive_stats}. \change{We found no significant differences in sound characteristics across time periods: bowel sounds occurring in the first and second halves of the WLT-II exhibited comparable acoustic properties (i.e., similar in duration, spectral centroid and spectral flatness). Bowel sounds occurred more frequently in the second half of recordings.}

\begin{figure*}[t]
    \centering
    \includegraphics[width=0.9\linewidth]{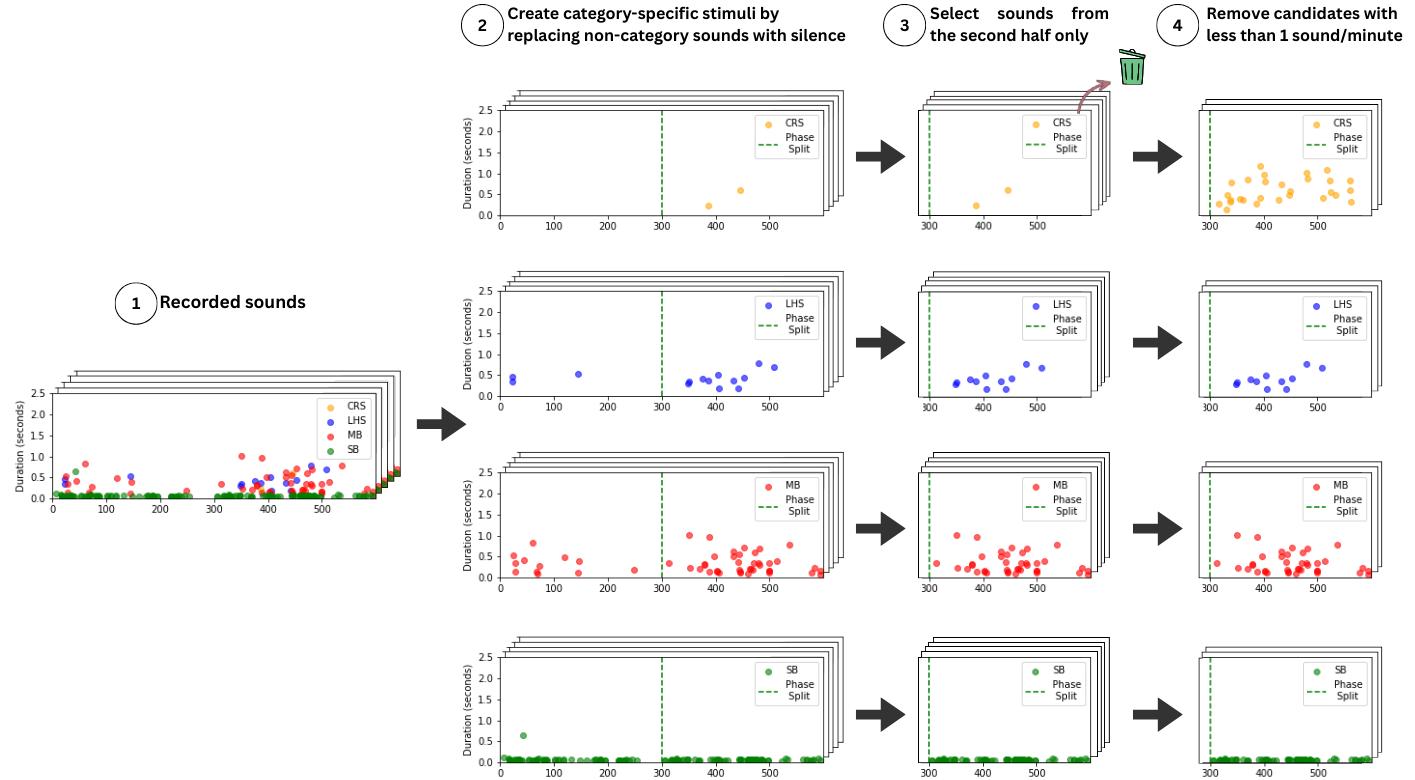}
    \caption{Stimuli selection pipeline. \Circled{1} Bowel sounds from 7 participants were recorded and labeled in two four categories. \Circled{2} Category-specific stimuli were created by isolating bowel sounds of each target category. \Circled{3} Keep the second half. \Circled{3} Remove candidates with less than 1 sound/minute.}
    \label{fig:stimuli_selection_pipeline}
    \Description{
     A Pipeline of stimuli selection with 4 steps: Stimuli selection pipeline. 1. Create pool: Bowel sounds from 7 participants were recorded and labeled in two four categories. 2: Create Category-specific stimuli: Isolating bowel sounds of each target category. 3. Remove the first half and keep the second half. 4. Remove candidates with less than 1 sound/minute.
    }
\end{figure*}

\textbf{Step 3: Stimulus pool curation}. We created category-specific stimuli by isolating bowel sounds of each target category and replacing sounds from other categories with silence, preserving the natural temporal patterns. This initially yielded 28 sound segments across 4 categories.
However, category frequencies varied dramatically—LHS and CRS each represented less than 5\% of total occurrences. Many segments contained insufficient sound instances (only 1-2 events lasting under 2.5 seconds across 10 minutes), particularly for LHS in participants P3 and P4 (Figure \ref{fig:bowel_sound_scattered}). 
\change{Table \ref{tab:phase_freq} shows the number of occurrences for each sound category per user. Bowel sounds has been found to have high individual differences \cite{du_bowel_2018}, and given the short recording time (10 minutes), it is expected that some sounds occurs more frequently in one person than in another person. Since we are interested in studying the effect of individual category, we aim to create a pool of stimuli in which each category of have several instances.}

\change{In the first 5 minutes, only two sound patterns had a total duration exceeding 10 seconds, compared to seven in the second 5 minutes}. Since bowel sounds occurred more frequently in the second half of recordings, we focused exclusively on the final 5 minutes to maximize the stimulation duration and at the same time reduce the potential temporal variance within categories. We excluded recordings with fewer than 5 sound instances (less than 1 per minute). 
This filtering process yielded 6 SB, 6 MB, 3 CRS, and 3 LHS segments. To ensure fair comparison, we selected 3 stimuli per category (matching the maximum available for CRS), creating a final pool of 12 stimuli.

\begin{table*}[h!]
\centering
\resizebox{\textwidth}{!}{
\color{\workingcolor}
\begin{tabular}{|l|rrrr|rrrr|rrrr|rrrr|}
\hline
& \multicolumn{8}{c|}{ {Phase 1}} & \multicolumn{8}{c|}{ {Phase 2}}\\
\cline{2-17}
& \multicolumn{4}{c|}{ {Total duration}} & \multicolumn{4}{c|}{ {Count}} & \multicolumn{4}{c|}{ {Total duration}} & \multicolumn{4}{c|}{ {Count}} \\
\hline
 {name} &  { CRS} &  { LHS} &  {MB} &  {SB} &  { CRS} &  { LHS} &  {MB} &  {SB} &  { CRS} &  { LHS} &  {MB} &  {SB} &  { CRS} &  { LHS} &  {MB} &  {SB} \\
\hline
P1 & 0.00 & 1.36 & 4.30 & 4.81 & 0 & 3 & 13 & 72 & \textbf{12.64} & \textbf{10.80} & \textbf{16.26} & \textbf{7.92} & 12 & 14 & 41 & 85  \\
P2 & 5.68 & 7.87 & 2.14 & 3.71 & 8 & 8 & 18 & 101 & \textbf{16.73} & \textbf{3.32} & 3.43 & 3.64 & 28 & 5 & 27 & 117  \\
P3 & 4.20 & 0.55 & 0.74 & 3.35 & 4 & 1 & 2 & 50  & 4.24 & 0.66 & \textbf{7.52} & \textbf{4.72} & 4 & 1 & 6 & 70  \\
P4 & 0.16 & 0.66 & 0.63 & 0.96 & 1 & 1 & 1 & 16  & 0.96 & 1.48 & 4.13 & 3.03 & 3 & 2 & 7 & 66  \\
P5 & 2.19 & 1.44 & 10.47 & 9.85 & 4 & 3 & 26 & 113  & \textbf{4.34} & 1.18 & \textbf{11.48} & \textbf{9.55} & 8 & 1 & 37 & 113  \\
P6 & 2.31 & 3.92 & 5.66 & 4.37 & 6 & 6 & 29 & 97 & 2.91 & 0.00 & 1.02 & 3.13 & 3 & 0 & 7 & 67  \\
P7 & 3.44 & 23.41 & 1.92 & 0.00 & 4 & 14 & 1 & 0  & 0.86 & \textbf{15.63} & 1.56 & 0.76 & 2 & 8 & 4 & 3  \\
\hline
\end{tabular}
 }
\caption[Number of sound instances per category per user]{\change{Number of sound instances per category per user. Sound instances are distributed across a 5-minute window for each phase, and total duration represents the sum of individual sound durations. Due to higher instance counts, Phase 2 was used for pattern selection. Within each category, the 3 patterns with both (i) at least 5 instances and (ii) highest total duration were selected as stimuli (bold).}}
\label{tab:phase_freq}

\end{table*}
\subsection{Participants}
18 participants (10 males, 8 females) aged 17-39 years (M=26.6, SD=6.2) took part in the study. \change{They were recruited from university pool}. All participants had eaten their last meal 1-2 hours before the session and received 7 USD compensation for their participation.  

\subsection{Apparatus}

We built the same set-up that were suggested by \citet{nguyen_gutio_2025} and used a chair (IKEA POANG \footnote{\url{https://www.ikea.com/sg/en/cat/poaeng-series-07472/}}) with two sound transducers (SolidDrive SD1\footnote{\url{https://soliddrive.mseaudio.com/sd-1sm-ti.html}}), one transducer was attached in the back and one can be freely moved. The free-moving transducer was placed on participant's stomach, in contact with their skin, right at their navel. This transducer was then fixed to participant's stomach using a back belt to limit movements when actuated. The system is connected to an amplifier and a Bluetooth receiver.
The apparatus is depicted in Figure~\ref{fig:setup_and_apparatus}. Note that the drinking setup was only used during our initial bowel sound recording session for stimulus creation and for our final study.
\begin{figure*}[t]
    \centering
    \subfigure[Number of reported instances of hunger and fullness per category]{
        \includegraphics[width=0.45\linewidth]{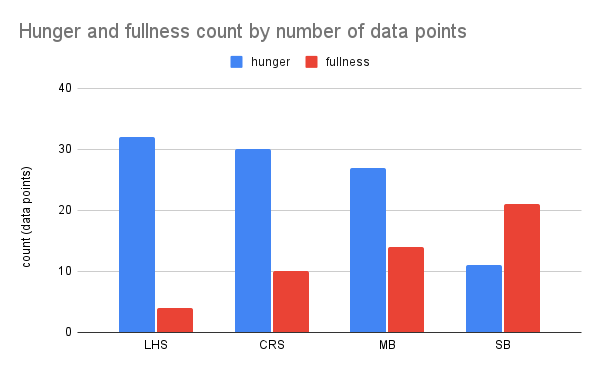}
        \label{fig:datapt_cnt}
    }
    \subfigure[Number of people experienced hunger and fullness per category]{
        \includegraphics[width=0.45\linewidth]{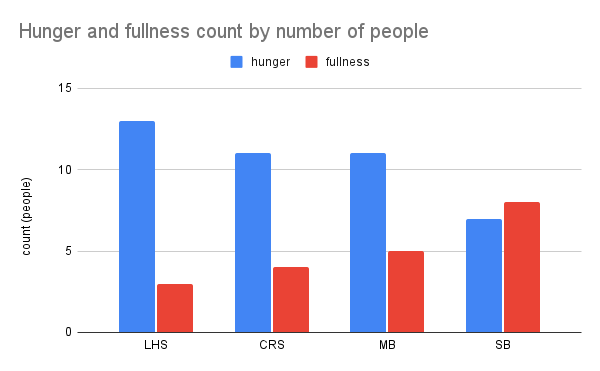}
        \label{fig:ppl_cnt}
    }
    \caption{Hunger and fullness analysis. }
    \label{fig:hunger_fullness_analysis}
    \Description{a: Bar chart showing "Hunger and fullness count by number of data points" across four conditions (lhs, crs, mb, sb). Blue bars represent hunger counts, red bars represent fullness counts. Hunger responses are highest for lhs (~32) and lowest for sb (~11), while fullness responses are highest for sb (~21) and lowest for lhs (~4).
    b: Bar chart showing "Hunger and fullness count by number of people" across the same four conditions. Blue bars show hunger reporting by participant count, red bars show fullness reporting. Pattern is similar to Image 1 but scaled to participant numbers rather than total data points, with lhs showing highest hunger reporting (~13 people) and sb showing most balanced hunger/fullness reporting (~7 people each).}
\end{figure*}
\subsection{Procedure}
We employed an open-ended thinking aloud procedure. Though we are interested in how the stimuli influence users' gastric feelings, we refrain from biasing them and ask them to openly state any feelings that they experienced. Upon arrival, participants were seated comfortably and listened to soothing music\footnote{https://youtu.be/f-i\_nJLG2Is?si=YDptfGa-tPuQLVoK}. We asked \textit{"how are you feeling?"} and waited for their response. Building on their answer, we explained: \textit{"You just experienced an auditory stimulus that made you feel calm/sad/nostalgic (depending on their answers). This demonstrates how stimuli can influence feelings. Over the next 20 minutes, you will experience 4 haptic stimuli, each lasting 5 minutes, which may or may not change your feelings. If you notice any change during the experience, please share your feelings with us. Otherwise, we will ask, `How are you feeling?' every minute. Feel free to tell us any feelings that you have on top of your head."}

Researchers then positioned the transducer on participants' abdomen directly over the navel and secured it with a waist belt. Participants were instructed to sit fully back against the chair to ensure contact with the transducer placed on the chair back.

We presented haptic stimuli individually using Latin-square counterbalanced category ordering, with random stimulus selection within each category. Participants reported their feelings upon sensing changes or responded to minute-by-minute prompts. \change{If a participant has already given a report on their own during a specific minute, we will not send them an additional prompt in that same minute.} We documented responses using a customized note-taking web application and audio recordings. After each stimulus session, participants provided brief reflections on their experience, concluding with an overall reflection after experiencing all patterns. 

\begin{figure*}[ht!]
    \centering
    \includegraphics[width=1\linewidth]{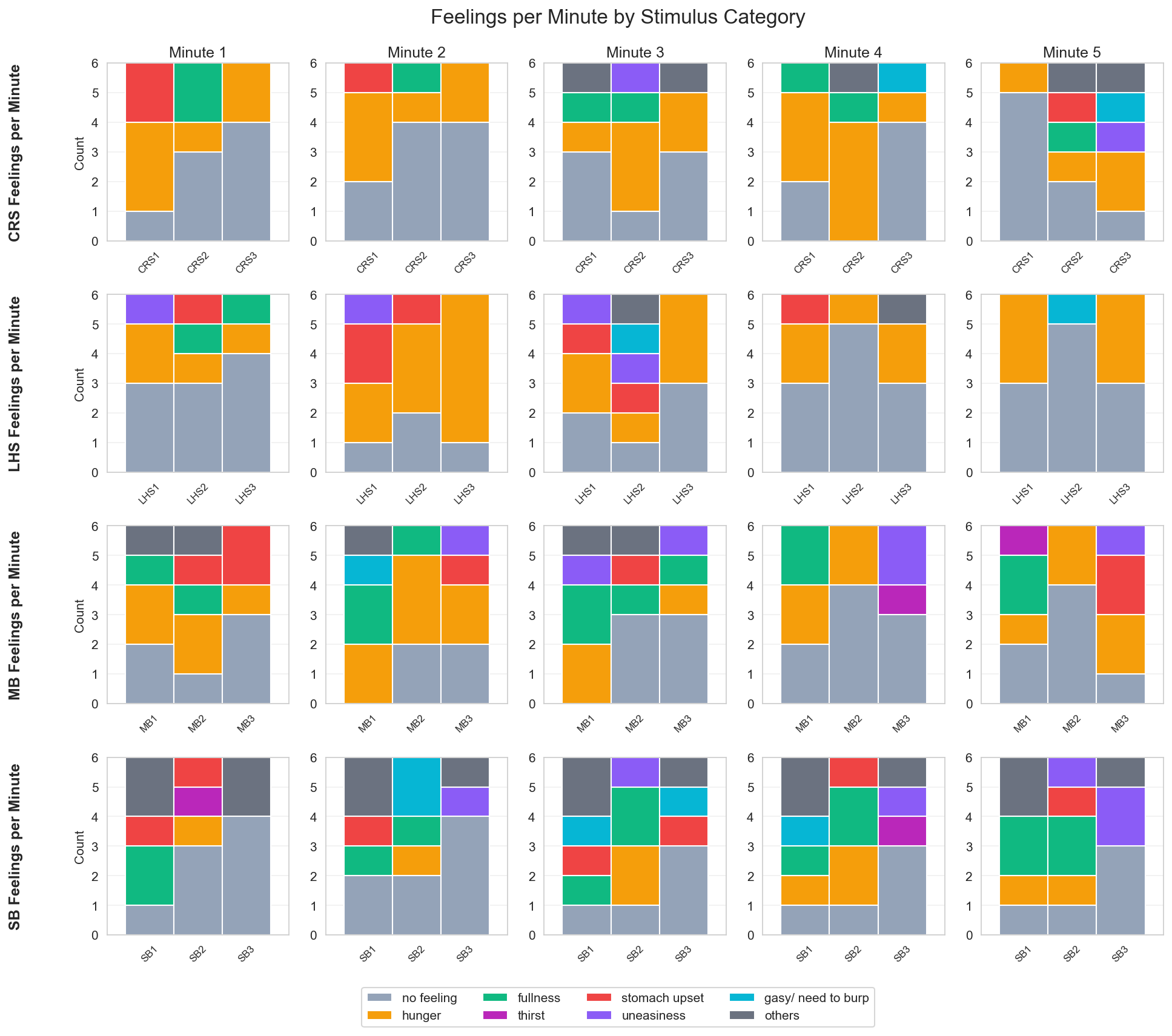}
    \caption{Minute-wise feeling distribution by individual stimulus. SB: Single Bust, MB: Multiple Bursts, LHS: Long Harmonic Sounds, CRS: Continuous Random Sounds.}
    \label{fig:by_stimulus}
    \Description{Stacked bar charts showing "Feelings per Minute by Stimulus Category" across four conditions (lhs, crs, mb, sb). Different colored segments represent various feelings like hunger, fullness, thirst, uneasiness, emotion, and physical sensations.
}
\Description{ Grid of stacked bar charts showing "Feelings per Minute by Stimulus Category" across 5 minutes for four different patterns (CRS, LHS, MB, SB). Different colored segments represent various feelings including hunger, fullness, thirst, stomach upset, and others}
\end{figure*}
\subsection{Data Analysis}
The lead author compiled and coded participants' minute-wise responses, which ranged from 1 to 43 words (M=6.96, Mdn=5, SD=6.36).  Following thematic analysis guidelines by  \citet{braun_using_2006}, the lead author familiarizes themself with the responses and iteratively groups them into themes. Unclear responses were discussed with the second author to either create a new theme or to put them into one of the existing themes.  

\subsection{Results}
Nine themes emerged from the coding process, namely: (1) no feeling, (2) emotional feelings, (3) physical feelings, (4) hunger, (5) fullness, (6) thirst, (7) stomach upset, (8) uneasiness, and (9) gassy/need to burp.

\textbf{Effects of gut-based audio-haptic feedback on gastric feelings:} Of the 174 gastric-related codes (48.3\% of total responses), we identified five categories: hunger (94 codes), fullness (34 codes), gassy/need to burp (12 codes), thirst (4 codes), and stomach upset (25 codes). Hunger emerged as the dominant feeling with straightforward expressions like \textit{"I feel hungry"} or \textit{"I feel like I need to eat something now."} Fullness responses were less frequent and more varied, ranging from direct statements (\textit{"I feel full," "I feel bloated/stuffed"}) to descriptive sensations (\textit{"my stomach feels like I've just had lunch"} or \textit{"my stomach feels like I've drunk a lot of water, and now the water is juggling"}).

Participants showed notable individual differences in interpreting bodily sensations. While some associated the bubbling sensations from SB and MB stimuli with fullness, others linked them to stomach upset and diarrhea, expressing concerns like \textit{"I feel like I need to go to the toilet"} or \textit{"I feel like I am going to have diarrhea."} This variability aligns with prior findings on individual differences in gastric cue interpretation. The gassy/need to burp category was uniquely observable, with actual burping recorded in 5 participants during the study.

To analyze the potential impact on appetite regulation, we focused on four key themes: hunger, fullness, thirst, and gassy/need to burp. We categorized these into second-order themes based on their likely effect on food intake—hunger and thirst as intake-encouraging feelings, and fullness and gassy/need to burp as intake-discouraging feelings. We quantified both the occurrence frequency of these themes per stimulus category (Figures \ref{fig:datapt_cnt}) and the number of participants experiencing hunger or fullness sensations for each category ( Figure \ref{fig:ppl_cnt}).

\change{Figure \ref{fig:datapt_cnt} showed a converging trend: the longer the sound, the more likely that it makes people feel hungry. LHS has the longest average duration per sound ($781\pm523 ms$), then CRS ($610\pm417ms$), followed by MB ($336\pm335ms$),  and SB has the shortest duration per sound ($63\pm43ms$). This is also noted by participants, with several participants stating that the longer the vibration, the hungrier they felt. For example, P1 mentioned \textit{"The longer vibrations make me feel hungry for the most part"}. Similarly while P5 said \textit{"when it's short vibration it felt like I'm digesting food, and when it's long vibration I feel like I'm hungry"}. }


\textbf{Effects on user's emotions and bodily sensations:} 
Since we did not specifically prompt for gastric feelings, participants spontaneously reported other experiences. Emotional responses accounted for 5.2\% of all reports, including nervousness, alertness, anticipation, disgust, and intrigue. Additionally, 3.8\% of reports described complex bodily sensations that participants struggled to categorize as discrete feelings. When unable to find precise words, participants offered vivid physical descriptions such as \textit{"I feel like there is a concert going on in my stomach"} or \textit{"I feel like I have just drunk a chunk of cold water."}

\textbf{\change{Quality of Induced Feelings:} }
Figure \ref{fig:by_stimulus} illustrates minute-wise feeling distribution by stimulus. Examining the rating of individual segments within a category, we can see strong variation within category: some segments are more likely to induce gastric feelings, while the other segments of the same category do not, as indicated by the dominance of the "no feeling" code in that segment.
 
In addition to stating specific feelings, participants also noted their varying intensities and how induced sensations interacted with their baseline states. Beyond simply \textit{"feeling hungry"} they described feeling \textit{"more hungry"} or \textit{"less hungry."} While these reports were all coded under the \textit{"hunger"} category, their effects on participants' feelings differed substantially. 
The dynamic nature of these changes within a short period of time is illustrated by P9's reflection during one stimulus segment: \textit{"I was feeling very hungry, not hungry at all, want to go toilet, and very hungry but less hungry (than the first time)."}. 


\change{\textbf{Study conclusions:}} This study demonstrated that gut-based audio-haptic feedback can induce gastric feelings, including hunger, fullness, stomach upset, and thirst. Closer examination of participants' experiences revealed substantial variation both between and within stimulus categories, with some patterns eliciting strong hunger or satiety responses while others produced minimal effects. Moreover, participants reported varying degrees of sensation intensity (e.g., \textit{"more hungry"} vs. \textit{"less hungry"}), and noted how these sensations interacted dynamically with their baseline physiological states.  

%

\begin{figure*}[htb]
    \centering
    \subfigure[Changes in hunger rating by group]{
    \includegraphics[width=0.44\linewidth]{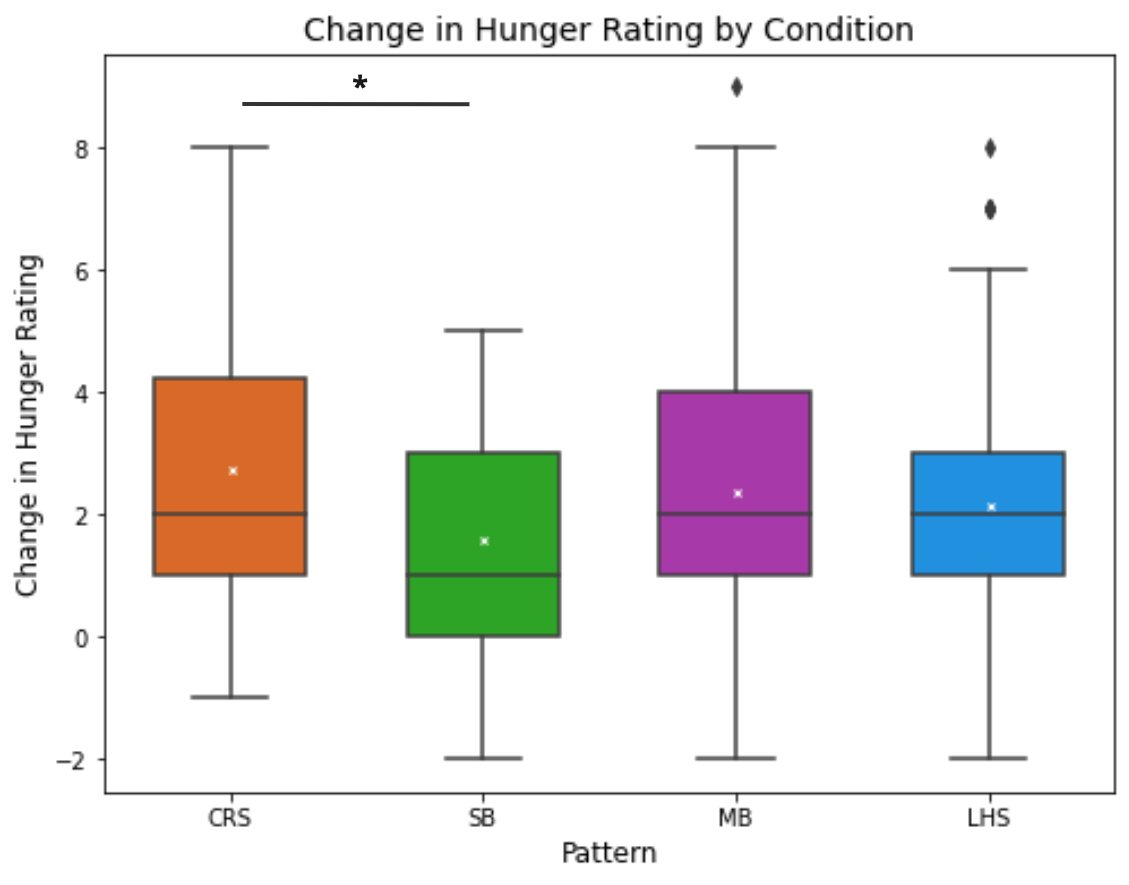}
    \label{fig:rating_box_plot}
   }
    \subfigure[Minute-wise change in hunger rating]{
    \includegraphics[width=0.45\linewidth]{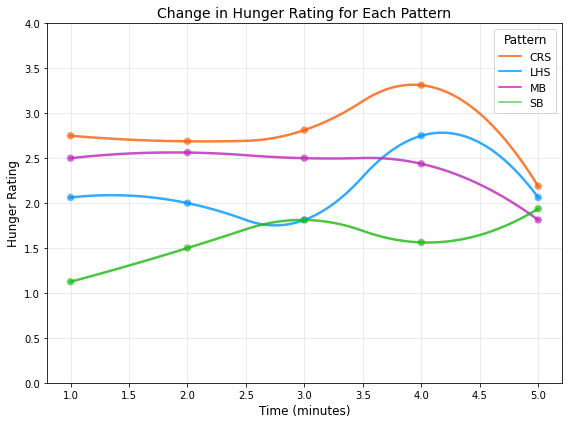}
    \label{fig:rating_change}
    }
    \caption{(a) Changes Hunger Rating Distribution by Conditon (b) \change{Average} Changes in Hunger Rating over Time. SB: Single Bust, MB: Multiple Bursts, LHS: Long Harmonic Sounds, CRS: Continuous Random Sounds }
    \Description{a: Box plot showing "Change in Hunger Rating by Condition" across four patterns (crs, sb, mb, lhs). Shows statistical significance (*) between crs and sb conditions, with crs showing higher hunger rating changes b. Line graph showing "Change in Hunger Rating for Each Pattern" over 5 minutes. Four colored lines represent different patterns (crs, lhs, mb, sb), with crs showing the highest peak around 3.5 minutes.}
\end{figure*}
\section{Study 2: Effects of Gut-based Audio-Haptic Feedback on Perception of Hunger and Satiety Level}

\begin{table*}
    
\centering
\resizebox{\textwidth}{!}{\begin{tabular}{lccccc}
\toprule
Group & \textit{n} & Change in Hunger Rating Mean & Change in Hunger Rating SD & Hunger Rating Mean & Hunger Rating SD \\
\midrule
CRS & 80 & 2.75 & 2.24 & 2.31 & 2.05  \\
LHS & 80 & 2.14 & 2.10 & 1.70 & 2.05  \\
MB  & 80 & 2.36 & 2.14 & 1.920 & 1.77  \\
SB  & 80 & 1.59 & 1.78 & 1.150 & 1.69  \\
\change{Baseline} & \change{16} & \change{-} & \change{-} & \change{-0.438} & \change{1.59} \\
\midrule
Total & 320 & 2.21 & 2.08 & 1.77 & 1.93  \\
\bottomrule
\end{tabular}
}
\caption{Descriptive Statistics of Hunger Rating and Changes in Hunger Rating by Group. \change{Baseline rating were taken when participants first come, while rating for CRS, LHS, MB, SB were taken every minute during a five minute stimulation session}.}


    \label{tab:study2_descriptive_stats}
    
\end{table*}
In Study 1 we observed that CRS and LHS sounds have a greater tendency to make people feel hungry, while SB sounds appear to make them feel full. Their reported minute-wise feelings, as well as their reflections at the end of each session, further revealed that the stimuli have varying effects on their hunger and fullness, with some making them feel hungrier than others.  

Quantifying how gut-based audio-haptic feedback influences the perception of hunger and fullness level adds further granularity to our understanding of the effect of gut-based haptic feedback on users' feelings. With our second study, we therefore want to answer the second research question: \textbf{RQ2:} \textit{"How does gut-based audio-haptic feedback influence users' perception of hunger and satiety level?"} 

\subsection{Stimuli Selection}
Figure \ref{fig:by_stimulus} revealed substantial within-category variance: stimulus segments within the same category varied significantly in their ability to influence users' feelings. Some segments had a negligible impact, with most participants reporting "no feelings." To \revise{maximize our chance of influencing perception of hunger and satiety levels}, we excluded low-impact segments and selected those with the highest probability of inducing gastric feelings. These selected segments were then combined to create one composite stimulus per experimental category. The selection criteria were:

\begin{itemize}
    \item Have the most number of people reported feeling hungry (for CRS and LHS) or feeling full (for MB and SB)
    \item When two segments have an equal reporting count, choose the segment that is less reported to create discomfort (i.e. stomach upset, uneasiness); whereby the prioritization order for discomfort was: reportings of stomach upset > uneasiness > feeling nothing.
\end{itemize}

\change{The selected segments are combined in the way that preserves their original temporal order (e.g. segment in minute 1 of stimuli SB1 will come before segment in minute 2 of SB2.)}

\subsection{Participants}
16 participants aged 17-32 years (M=24.25, SD=4.78) \change{were recruited from the university's pool and the local community}, with no overlap from Study 1. The sample included 12 males and 4 females. Participants self-rated their awareness of gastric activities as Poor (n=2), Average (n=8), or Good (n=6). \change{They were compensated 7 USD for their time.}

\subsection{Apparatus and Procedure}
We used the same apparatus setup as in Study 1.
Participants followed a similar procedure in Study 1. Additionally, every minute, they were asked to rate their hunger and satiety level on a scale of -7 to 7, with -7 being extremely full, 0 being neither full nor hungry, and 7 being extremely hungry. \change{The scale is adopted from the bipolar Satiety Labeled Intensity Magnitude (SLIM) scale developed by\cite{cardello_development_2005}}. Participants also reported their initial hunger level before they were exposed to any haptic stimulus. 

\subsection{Data Analysis}

\textbf{Hunger/Satiety Rating:}
Changes in hunger rating were calculated by subtracting the hunger rating from the initial hunger rating. 
To determine main effect of stimulus category to changes in hunger level, we use Repeated Measure ANOVA. Shapiro tests are used to determine normality. When normality is not met,  Aligned Rank Transform (ART) ANOVA \cite{wobbrock_aligned_2011} for repeated measures is used. Post-hoc pairwise comparison with Holm correction is then used to determine pairwise differences. 

\textbf{Subjective Feelings:} Similar to study 1, we coded users' reports of subjective feelings every minute using iterative coding.


\subsection{Results}

\textbf{Hunger Ratings:}
Descriptive statistics of hunger and satiety level ratings are reported in Table \ref{tab:study2_descriptive_stats}. Average changes by minutes are plotted in figure \ref{fig:rating_change}. Overall, the mean hunger rating change was 2.21 (with a standard deviation of 2.08), and the median value was 2.00. When comparing the groups, the CRS group demonstrated the highest mean change at 2.75 (SD=2.24), while the SB group showed the lowest average change at 1.59 (SD=1.78). The LHS and MB groups had similar mean hunger rating changes of 2.14 (SD=2.10) and 2.36 (SD=2.14), respectively. The median hunger rating change was 2.00 for all groups except for SB, which had a median of 1.00

Normality testing revealed significant violations of the normality assumption across all groups (all p < 0.05). The overall dataset also failed the normality test (p < 0.001), therefore Aligned Rank Transform (ART) ANOVA was performed to determine main effect of category on changes in hunger rating. The ART ANOVA revealed a statistically significant effect of group on changes in hunger ratings, (\textbf{$p = 0.018, \eta^2 = 0.031$}). ART post-hoc pairwise comparisons using Holm correction revealed that only the contrast between CRS and SB groups was statistically significant (\textbf{$p = 0.015$}). Specifically, the CRS group showed significantly higher hunger rating changes compared to the SB group.
\begin{figure*}[t]
    \centering
    \includegraphics[width=1\linewidth]{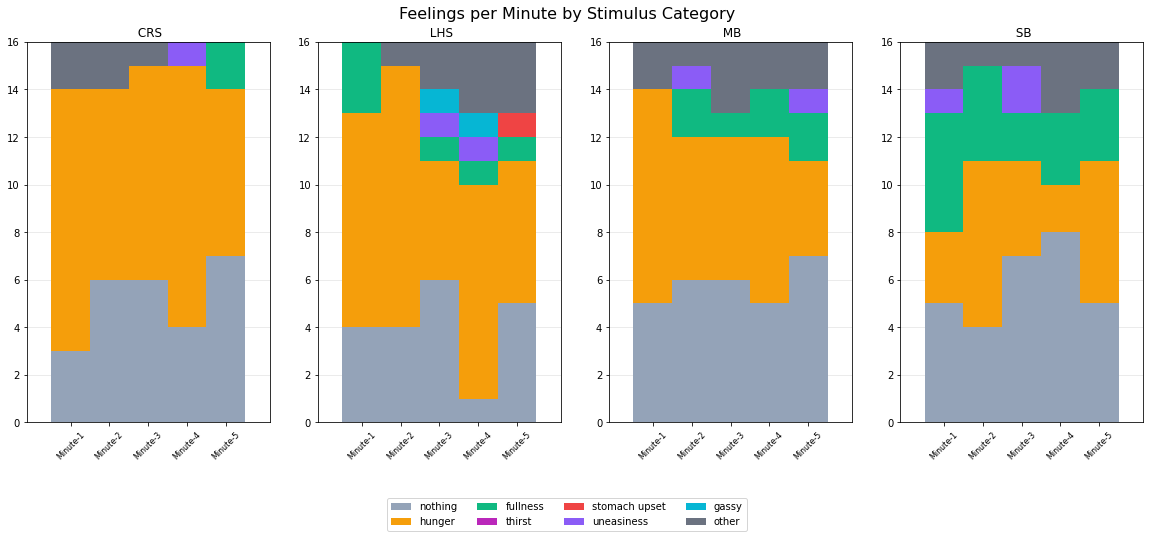}
    \caption{Feeling rating by Minute by Stimulus Category. SB: Single Bust, MB: Multiple Bursts, LHS: Long Harmonic Sounds, CRS: Continuous Random Sounds}
    \label{fig:study2_qual}
    \Description{Grid of stacked bar charts showing "Feelings per Minute by Stimulus Category" across 5 minutes for four different patterns (CRS, LHS, MB, SB). Different colored segments represent various feelings including hunger, fullness, thirst, stomach upset, and others.}
\end{figure*}

\change{\textbf{Subjective Feelings:}} Figure \ref{fig:study2_qual} showed minute-wise feeling distribution. Consistent with the finding from Study 1, hunger is the most felt feeling under the influence of gut-based audio-haptic feedback. We continue to observe within-category variances, depicted in the fluctuation of changes in hunger rating  over time (Figure \ref{fig:rating_change}). Minute-wise rating also revealed an interesting detail not observed before: although LHS continued to receive a high number of reports on minute-wise hunger feelings (depicted in the number of people who said they felt hungry experiencing the stimuli, Figure \ref{fig:study2_qual}), their hunger ratings in the first three minutes were only higher than the rating of SB (Figure \ref{fig:rating_change}). Notably, in the second minute, 11 people reported experiencing hunger during LHS, compared to 6 experienced hunger during MB. Yet, the average hunger rating of MB of that minute is higher than that of LHS. This discrepancy highlights that gut-based haptic feedback affects not simply whether one feels hungry but the degree of hunger experienced.

\change{
\textbf{Study conclusions:} This study quantified the changes in perceived hunger level created by gut-based haptic feedback, providing a more granular lens to the quality of users' experience. We found that (1) CRS stimulus make people felt significantly hungrier compared to SB stimulus and (2) within-category sub-segments have varying effects on users' hunger level.
}

\section{Study 3: Effect of Gut-Based Audio-Haptic Feedback on Gastric Interoceptive Behavior}

Studies 1 and 2 combined demonstrated that gut-based audio-haptic feedback induced gastric feelings in participants, evidenced by both momentary self-reports and quantitative hunger ratings. Verbal reports from participants further indicated that gut-based audio-haptic feedback may influence their behavior. Hence, we conducted a third study which aimed to address \textbf{RQ3}: \textit{"Is it possible to non-invasively modulate gastric interoceptive behavior through gut-based audio-haptic feedback?"}

\subsection{Stimuli Selection}
Previous studies revealed that segments within categories have varying effects on hunger. When combined, weaker segments may dilute the impact of more potent ones, reducing overall effectiveness. To establish whether non-invasive influence of gastric interoceptive behaviors is possible, we sought to avoid this averaging effect, which could reduce effect sizes below our detection threshold and yield false negatives. Therefore, we selected two contrastive segments with demonstrated strong effects: CRS from minute 4 and SB from minute 1. \change{To stimulate participants, the selected 1-minute segments were continuously replayed throughout the stimulation period. }

\subsection{Participants}
Participants were recruited from the university pool. Exclusion criteria being (1) having known digestive problems, (2) being pregnant (3) on psychological medication in the last 2 months. 21 participants took part in the study (10 males, 11 females, age range =\change{ 18 to 34 , mean=24}; 9 Native/Bilingual English speakers, 12 Professional-level Enghlish speakers). \change{No participants participated in the previous studies. They were compensated 25 USD for their time}

\begin{table*}[h]
\centering
\begin{tabular}{l|cc|cc|cc|cc}
\hline
& \multicolumn{2}{c|}{sat\_ml} & \multicolumn{2}{c|}{full\_ml} & \multicolumn{2}{c|}{total\_ml} & \multicolumn{2}{c}{sat\_\%} \\
\cline{2-9}
Type & Mean & Std & Mean & Std & Mean & Std & Mean & Std \\
\hline
CRS & 245.476 & 127.990 & 347.810 & 192.793 & 593.286 & 276.104 & 41.4 & 9.7 \\
SB & 294.524 & 148.717 & 328.476 & 199.505 & 623.000 & 291.124 & 47.1 & 14.8 \\
Baseline & 249.190 & 179.443 & 325.714 & 170.852 & 574.905 & 264.266 & 41.9 & 15.0 \\
\hline
\end{tabular}
\caption{Statistical Summary Ingested Volumes During WLT-II}
\label{tab:wlt-stats}
\end{table*}

\subsection{Apparatus}
To deliver haptic feedback, we used the same setting as Study 1 and Study 2. For the WLT-II, 2-liter bottles filled with room-temperature water is used. To replicate the test in the literature, we cover the bottle with a dark-colored bottle sleeve. 1-meter long straws were used to control drinking speed, as well as to ensure participants sit in a stable position while drinking the water. Figure \ref{fig:setup_and_apparatus} illustrates the setup of the WLT-II.

\subsection{Measurements}
\subsubsection{Gastric Interoception}
We used the WLT-II (described in \ref{section:wlt}) to quantify participants' gastric interoceptive behavior. We recorded the volumes of ingested water during each drinking phase and named them as follows:
\begin{itemize}
    \item sat\_ml: the water volume ingested in the first 5 minutes (Drink until satiation phase, or satiation phase for short)
    \item full\_ml: the water volume ingested in the second 5 minutes
total (Drink until full phase, or full phase for short)
    \item total\_ml: the total volume of water ingested in both phases
    \item sat\_\%: Satiation to total volume ratio, calculated by the following formula: $sat\_\% = sat\_ml/ total\_ml$
\end{itemize}

\subsubsection{Affect and Fullness Perception}
\label{section:affect}
\change{Following the measurements protocol by \cite{van_dyck_water_2016}}, momentary self-rating of perceived fullness and negative feelings such as thirst, stomach tension, immobility, discomfort, guilt, sluggishness, nausea, and arousal on a Likert scale of 1 (not at all) to 7 (extremely) were used to capture participants' affective states before the WTL-II test (t0), after the first (t1) and the second drinking period (t2).

\subsection{Procedure}

\change{We employed a within-subjects design with participants completing the Water Load Test (WLT-II) across three sessions on separate days. Each session featured one condition: \textbf{baseline (no stimulus)}, \textbf{SB stimulus}, or \textbf{CRS stimulus}}. Since participants need to come over multiple sessions, we informed participants in advance that there would be two drinking sessions to keep the information consistent across all sessions. This follows the recommendation of \citet{van_dyck_water_2016}. In each session, they experienced one of the comparing conditions. The procedure for one session is as follows:
\begin{enumerate}
    \item \change{\textbf{Prerequisite:}} Participants were asked to fast for 3 hours and refrain from any liquid drinks for 2 hours. When they signed up, the research team coordinated their sessions so that all their sessions were scheduled on the same time slot of different days. Information about the study was given to them in the first session, and written consent was taken.
    \item \change{ \textbf{Pre-study:}} Before the study started, participants were asked to meditate for 5 minutes using a guided meditation audio to minimize the potential effect of stress. Once they finished the meditation session, they were asked to fill in a questionnaire that assesses their initial state of fullness, thirst, and other affective states described in section \ref{section:affect}.
    
    \item \change{\textbf{Set up:}} The researchers then introduced the haptic device to them and helped them put the transducer on their stomach, right on the top of their navel. The transducer position is secured with a waist belt and was tightened to avoid creating pressure on the participant's stomach. 

    \item \change{\textbf{Calibration:}} The participants were then asked to sit comfortably on the haptic chair, leaning all the way against the back of the chair so that the second transducer placed hidden in the back could contact their body. The researchers then perform calibration to ensure that the amplitude of the \change{stimulus} are (1) not uncomfortable \change{and} (2) realistic, as if it \change{comes} from their own body. To maintain the consistency between sessions, both of the stimuli were calibrated regardless of the session type. The initial weight of the bottle was measured using a kitchen scale and was noted down \change{on} a \change{Excel} sheet. 

    \item \change{\textbf{Water Load Test:}} After the calibration, researchers explained the water load test procedure (described in section \ref{section:wlt}). They were also told to sit relaxed in the chair if they finished drinking before the 5 minutes ended. After 5 minutes, the researcher measured again the weight of the water bottle, reminded the participant to fill in the affective questionnaire, and explained the second phase of the water load test to them. After another five minutes, the researcher repeated the procedure of the first half of the WLT-II. 
    
    \item \change{\textbf{Post-study Interview:}} Once the participants finish answering the affective questionnaire, the researchers interview them for 5 minutes to understand their experience, whether they feel any sensations, and how they decided when they should keep drinking and when they should stop drinking. 
\end{enumerate}
\subsection{Data Analysis}

 Aside from the key indices in WLT-II, we also computed the difference between the measure obtained for the sessions in which the feedback was applied, and the measure
obtained for the baseline (no feedback). We indicated these measures with \texttt{\_change} suffix.   Shapiro-Wilk tests were used to check normality assumptions. When normality assumptions were violated, the Aligned Rank Transform was then used to transform the data before applying Repeated Measure ANOVA. Post-hoc pairwise comparison with Holm correction was used to determine significance between conditions. For comparing the changes between the CRS and SB conditions, we used paired student T-tests when the normality assumption was met and Wilcoxon signed-rank test when the normality assumption was violated.

\begin{figure*}
    \centering
    \includegraphics[width=1\linewidth]{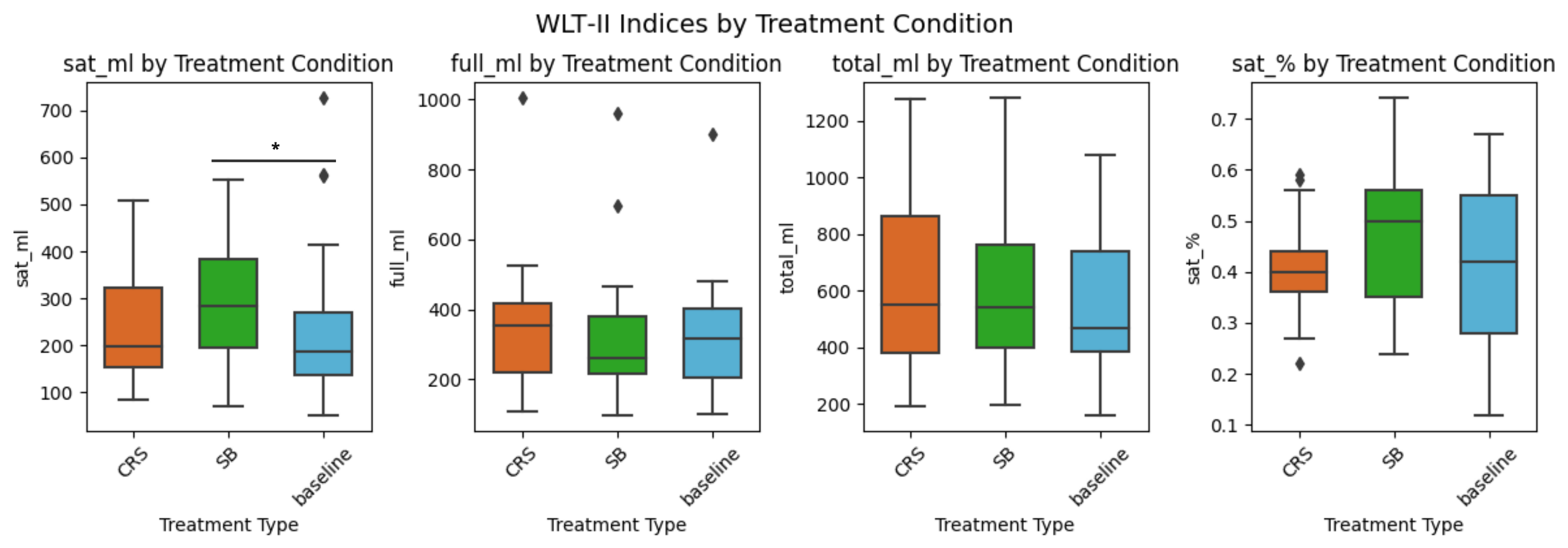}
    \caption{Volumes of ingested water by condition. SB: Single Bust, MB: Multiple Bursts, LHS: Long Harmonic Sounds, CRS: Continuous Random Sounds. $^*p<0.05$}
    \label{fig:wlt-ii}
    \Description{ Box plots showing WLT-II indices (sat_ml, full_ml, total_ml, sat_\%) across three treatment conditions (CRS, SB, baseline). Shows statistical significance marker (*) between SB and baseline for sat_ml.
}
\end{figure*}

\begin{figure*}
    \centering
    \includegraphics[width=1\linewidth]{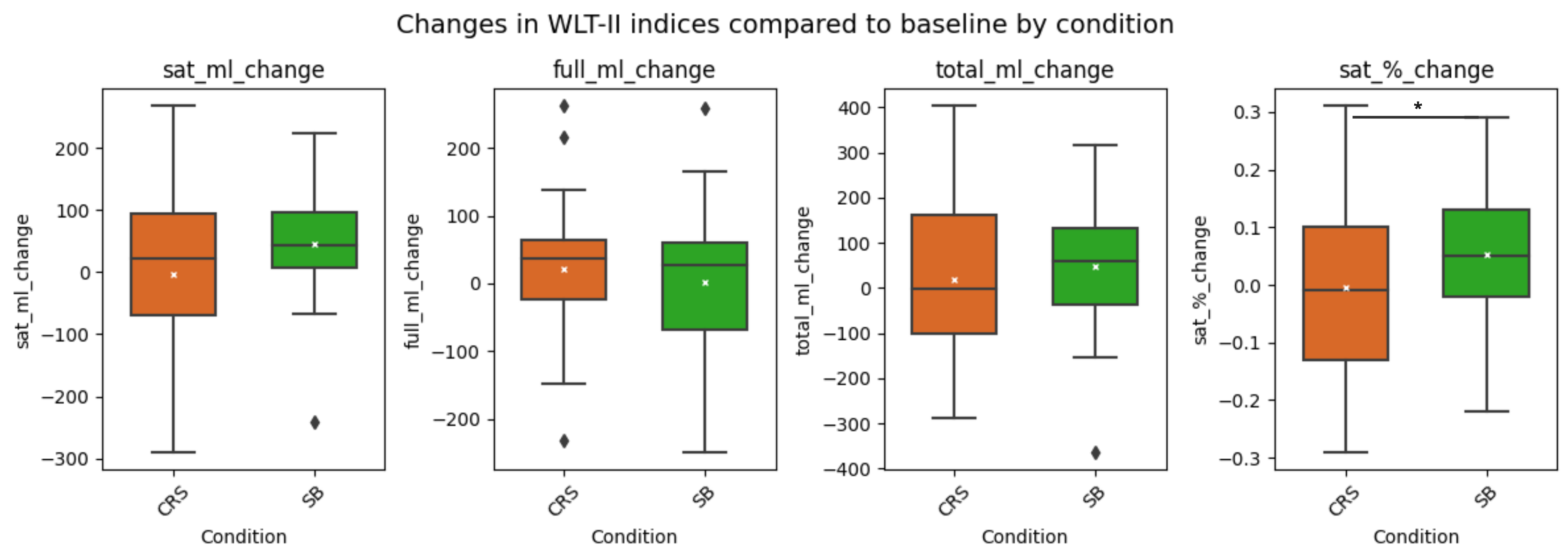}
    \caption{Changes in volume ingested compared to baseline by condition. SB: Single Bust, MB: Multiple Bursts, LHS: Long Harmonic Sounds, CRS: Continuous Random Sounds. $^*p<0.05$}
    \label{fig:changes-in-wlt}
    \Description{Box plots displaying changes in WLT-II indices compared to baseline for CRS and SB conditions. Shows statistical significance (*) for sat_\% change between conditions
}
\end{figure*}

\subsection{Results}
A total of \change{63} sessions were recorded. Table \ref{tab:wlt-stats} summarized the descriptive statistics of water drunk during the satiation phase (sat\_ml), the full phase (full\_ml) and the total amount of water drunk (total\_ml). 

\textbf{Effects of gut-based audio-haptic feedback on gastric interoception indices:} 
The normality assumption was not met for satiation volume (sat\_ml) satiation to full volume (full\_ml), and total volume (total\_ml) was not met ($p<0.05$) but was met in gastric sensitivity ($p>0.05$). Therefore, ART-ANOVA was used to determine the main group effect for sat\_ml, full\_ml, total\_ml and Repeated Measure ANOVA was used for sat\_pct. 

\textbf{Satiation Volume:} A significant main effect was found on satiation volume (\textbf{$p=\textbf{0.044}$}). Post-hoc pairwise comparison with Holm correction showed a significant difference between SB and baseline conditions (\textbf{$p=\textbf{0.049}$}), with participants drinking significantly more water in the SB condition than in the baseline condition. No significant difference was detected in other pairs ($p>0.05$)

\textbf{Full Volume and Total Volume: }No significant main effect was found on satiation to full volume ($p=0.0516$), total volume ($p=0.531$) or gastric interoceptive sensitivity ($p=0.309$). Figure \ref{fig:wlt-ii} shows the box plots for the various gastric interoception indices between groups. 

\textbf{Changes in gastric interoception indices compared to baseline:} 
Normality assumptions on satiation volume change (sat\_ml\_change) and total volume change (total\_ml\_change) were not met
but satiation to full volume change (sat\_to\_full\_ml\_change),  and
(sat\_pct\_change) was met ($p>0.05$), so paired t-tests were performed to determine the differences in sat\_pct\_change and full\_ml\_change and Wilcoxon signed-rank tests were used to determine the differences in sat\_ml\_change and total\_ml\_change. No significant differences were found on sat\_ml\_change ($p=0.305$), full\_ml\_change ($p=0.393$) or total\_ml\_change ($p=0.903$). However, significant differences were detected in sat\_pct\_change ($p=\textbf{0.044}$), in which the SB stimuli increased the satiation percentage by 5\% while the CRS stimuli did not change the satiation percentage. Figure \ref{fig:changes-in-wlt} shows the box plots for changes compared to baseline of these various gastric interoception indices.

\textbf{Affective Rating:}
Since affective ratings were collected at 3 different time points (before drinking (t0), after the first 5-minute (t1) and after the whole drinking session (t2)), we investigated the influence of haptic feedback on affective ratings at each time point by analyzing the main effect of haptic condition on affective rating at t0, t1, and t2 respectively. e.g., at t0, we compared the affective rating in SB, CRS, and baseline conditions.

No significant main effect of conditions on any affective rating of "fullness", "sluggish", "guilt", "discomfort", "nausea" at any given time point. "stimulated" rating did not show any significant main effect at t0 and t1 but showed a significant main effect at t2 ($p=0.163$), with post-hoc tests with Holm correction revealing a significant difference in rating between baseline and CRS ($p=0.02$). However, the average rating for stimulated at t2 remains low. 

\change{\textbf{Results consolidation:}} Together with other gastric interoception indices, the quantitative results showed that participants under the SB condition drank significantly more water during the satiation phase compared to the baseline condition. This suggests that during the satiation phase, participants felt hungrier in the SB condition and needed more water to fill their stomach to feel satisfied. This interpretation is supported by the fullness ratings: despite drinking significantly more water in the SB condition, participants reported similar fullness levels across conditions. 


\change{\textbf{Study conclusions:} This study investigated the possibility of non-invasively manipulating gastric interoceptive behaviors with gut-based audio-haptic feedback. The significant increase in water volume consumed during the satiation phase under the SB condition, along with significant changes in the satiation-to-full ratio, suggests that gastric interoception can indeed be influenced through audio-haptic feedback. To our knowledge, this represents the first demonstration of non-invasive gastric interoception modulation.}
\section{Discussion}


\subsection{Summary of Findings}
\subsubsection*{RQ1: How Gut-based Audio-Haptic Feedback Influences Feelings in the Here and Now}

\change{Study 1 employed a think-aloud protocol to investigate what feelings are invoked by gut-based audio-haptic feedback. The open-ended nature of this protocol allowed us to capture the feelings that naturally arise as the result of the stimuli. The results suggest that gastric feelings can be induced by gut-based audio-haptic feedback. The most dominant feelings reported were hunger ($55.6\%$), followed by fullness ($20.1\%$), and stomach upset ($14.8\%$). These feelings translate to changes in hunger perception which we observed in study 2. 
Most importantly, the observed changes in feelings were strong enough to actually influence participants' behavior, reflecting in a significant increase in the volume of ingested volume seen in study 3.}

\change{Aside from gastric feelings, emotions such as nervous, disgust, intrigued or anticipation were also reported, albeit in small number. Modern affective science maintains that emotions are context-dependent \cite{barrett_affect_2009, barrett_conceptual_2014}, e.g., a churning stomach before lunch can mean hunger, but the same sensations can be interpreted as "nervousness" if it happens before an exam \cite{barrett_how_2018}. Our study did not take context into account, which could have resulted in fewer reports of emotional feelings. Though limited, the few reports on emotions suggest that gut sensations can also be linked with certain emotions such as nervous, disgust or anticipation. Future research can further investigate if gut-based sensations can reliably evoke emotions when additional context is provided during stimulation.}

\subsubsection*{RQ2: How Gut-based Audio-Haptic Feedback Influences Perceived Hunger Level}
\change{Study 2 investigated how different audio-haptic patterns influence hunger levels. Participants reported significant increases in hunger compared to baseline, with notable pattern-specific effects: CRS increased hunger ratings more substantially than SB in minute-wise comparisons.}

\change{Triangulating quantitative ratings with qualitative reports revealed a nuanced distinction between breadth and depth of hunger induction. LHS consistently elicited hunger sensations across most participants, yet produced smaller magnitude changes than CRS and MB during the first three minutes (Figure \ref{fig:rating_change}). Conversely, while fewer participants reported feeling hungry under the influence of  MB, those who did experience comparable hunger level changes to other high-intensity patterns. This discrepancy reveals two distinct dimensions of the stimuli's quality: some patterns achieve broad reach with consistent but mild hunger cues, while others produce pronounced responses in users. These findings demonstrate that gut-based stimuli have diverging effects on users' feelings. Consequently, inducing specific hunger sensations requires carefully designed stimuli.}

\subsubsection*{RQ3: Altering Gastric Interoceptive Behavior Through Non-invasive
Audio-haptic Feedback}
%

 %
\change{The results from study 3 showed that during the satiation phase, participants drank more water in the SB condition compared to baseline, where participants did not receive any feedback.  When taking the baseline as the ground truth, we observed that changes in the satiation-to-full ratio increased significantly in the SB condition compared to the CRS condition. These findings suggest two key implications: (1) Gastric interoceptive behaviors can be influenced noninvasively through manipulating perceived gastric feelings; (2) this effect is stimulus-specific, as the CRS stimuli did not produce the same result. This could open up new opportunities for future work to design applications that involve nudging users' gastric interoceptive behaviors (such as eating-related applications). }

\subsection{Designing Gut-based Audio-Haptic Feedback to Induce Gastric Feelings: Insights and Reflections}


\change{Qualitative results revealed an important insight: stimulus from the same-category showed varying impacts on feelings. Some segments are more likely to induce gastric feelings, while the other segments of the same category do not, as indicated by the dominance of the "no feeling" code in that segment (cf. Figure \ref{fig:by_stimulus}).} This within-category variation is also observed by participants, with participants noting how the varying characteristics of the individual sounds, as well as how often they occur, induced certain gastric feelings. \change{To inform future research, we consolidate key factors that were reported to influence users' feelings as follows:}

\subsubsection*{Duration of a single sound instance: }
Figure \ref{fig:ppl_cnt} illustrates that longer sounds (such as those in LHS) induced the feeling of hunger more often than shorter ones. Figure \ref{fig:datapt_cnt} suggests a trend: as the average sound duration decreases, the number of `hungry' reports also decreases. \change{Participants' reflections provide a potential explanation for this phenomenon: the sequence of short bursts create the bubbling sensations, while the longer sounds, were linked with hunger, as audible growling sounds that are often heard when the stomach is empty. For example, P1 mentioned \textit{"The longer vibrations make me feel hungry for the most part"}. This explanation aligns with the wider literature on gastric interoceptive cues, where the presence of stomach growls and stomach pains were associated with hunger\cite{murray_Consumer_2009}.}

\subsubsection*{Amplitude of Sound: }
\change{Participants report how the amplitude (loudness) of sound segments also influence the types of feelings induced. More concretely, several participants said that for LHS sounds, the \textit{"stronger intensity sounds made me feel like I needed to go to the toilet, but the weaker intensity ones made me feel hungry"} (P10). Similarly, P9 commented on MB sound: \textit{"The loud bumps make me feel like I need to go to the washroom. when there's heavier vibration, it makes me feel like my stomach is upset; otherwise, it feels normal."}}

\change{Aside from impacting \textbf{what} type of gastric feelings the stimuli induced, sound amplitude can also influence \textbf{whether} gastric feelings were induced at all. Participants noted that when the sounds get too loud, they tend to feel them as external which could explain the ``no feelings" reports: \textit{"The stronger ones felt a bit external"} (P7, P13).} 

\subsubsection*{Personal experiences: } \change{The familiarity of the sensations created by the stimuli also influences whether gastric feelings were induced. Participants state that when some sensations were perceived to be unfamiliar, they could not associate them with any feelings (and hence reported no feelings). \textit{"I don't really get the pops. The pops were quite strong, but the unrealistic part was because I don't [normally] get those pops [naturally]"} (P15). }
\vspace{-0.2cm}
\change{\subsubsection*{Saliency of bodily sensations.} The results of Study 3 showed no significant differences in the volumes of ingested water during the second phase. Interviews with participants suggest that as participants consumed more water, their natural satiation signals intensified, creating a noticeable discrepancy with device-induced sensations. P8 explained: \textit{"Initially, the feedback could trick me into feeling hungry and made me drink more water; however, as my stomach filled, the feedback could not trick me anymore."} Most participants reported relying on physical discomfort cues (nausea, stomach distention) to determine when to stop drinking in later phases}.

{
\color{\workingcolor}
\subsubsection*{Occurrence Frequency} Several sound segments failed to induce feelings. Participants linked this to the infrequent occurrence of shorter-duration sounds. For instance, within CRS, P14 reported that \textit{"the short vibration isn't really felt much, but the longer ones made me feel hungry"}. This result is intuitive, because the durations of CRS and LHS range from $128\text{ ms}$ to $2.5\text{ s}$, and they occur far less frequently than MB and SB (CRS and LHS occur less than 6\% of the time). Consequently, a brief burst lasting less than a fraction of a second is unlikely to induce any noticeable feelings.

In contrast, within Study 3, CRS did not significantly influence the amount of water drunk. This results is counterintuitive, as CRS was rated to induce higher level of hunger in Study 2. Interviews with participants suggests that the too-frequent occurrence of CRS appeared to heighten participants' awareness, making them more cautious. This subsequently could have reduced the effectiveness of CRS in influencing gastric interoceptive behaviors.  Several participants indeed remarked that \textit{"it vibrates, and it kind of makes me more aware of my stomach"} (P20) and \textit{"I think the vibration made me more aware of my stomach, so I paid more attention to whether I'm hungry or full"} (P5). This dual observation highlights the challenges when designing audio-haptic feedback to influence users' behaviors: sounds that are too infrequent fail to exert influence, while sounds that are too frequent may inadvertently make users overly cautious.
}

\subsubsection*{Potential Working Mechanism: } 

\change{The reflections of participants on how factors such as duration, loudness, and personal experience influence the type of gastric feelings induced, align with predictive coding theories \cite{friston_free-energy_2010}. These theories propose that the brain infers meaning of sensory signals by comparing them against learned models and minimizing prediction errors \cite{friston_free-energy_2010, seth_active_2016, sterling_allostasis_2012, barrett_how_2018}. Since natural bowel sounds are perceived through both auditory and tactile channels, external audio-haptic stimuli replicate this experience and integrate into the brain's predictive model, enhancing perceived realism. This realism was evident across all three studies: participants frequently reported that tactile sensations felt as if they originated from their own body rather than an external device. As P9 noted, \textit{"I was trying to separate the growling of my stomach and the sensations, and I don't know if the growling was induced by the haptic [device] or not"} (Study 1).
}

\change{As the brain sees external signals as internal, it assigns the corresponding meaning to the sensations using past experience \cite{barrett_how_2018}. This explains why people tend to associate long sounds with hunger, loud sounds with stomach upset (reflecting the GI tract's hyperactivity when expelling toxins), and only report gastric feelings when the sensations align with their prior experience. In contrast, when the signals from external stimulus conflict with the internal signals, the discrepancy is brought up to conscious awareness and the brain deems the external signals as irrelevant.}

\change{
After assigning meaning to sensations, the brain generates predictions that trigger bodily changes \cite{schulkin_allostasis_2019}. Participants reported: \textit{"[I] felt changes in the bodily feeling. [I] feel like I'm super hungry and need to run [to eat] to sort out my stomach problem."} (P1, Study 1); \textit{"I do crave for a bar of chocolate now."} (P6, Study 1). These findings align with other interoceptive studies documenting physiological changes during interoception manipulation \cite{costa_boostmeup_2019, valente_modulating_2024}.}

\subsection{Implications}


\subsubsection*{Gastric sensations as a new modality for interaction: }


Findings from  study 1 \& 2  corroborate the preliminary exploration by \cite{nguyen_gutio_2025} and demonstrate that visceral sensations, particularly gastric sensations, can be reliably augmented through haptic feedback.
This capability opens new possibilities for incorporating gut sensations as a design element in \change{human-computer interfaces. For instance, gut sensations can be integrated into existing educational games such as Go-Go Biome\cite{pasumarthy_go-go_2024} or Goey Gut Trail \cite{pasumarthy_gooey_2021} to enhance playfulness and immersion. Researchers and designers could also harness gastric sensations as behavioral cues in immersive games \cite{brooks_augmented_2024}, mindful-eating application \cite{sas_opportunities_2025} or use gut feelings as a way to give the gastrointestinal tract a "voice," facilitating novel forms of interoceptive interaction with one's own body.}

\change{The HCI and haptic community has a long-standing interest in creating haptic feedback to communicate feeling over distance\cite{wang_keep_2012,yang_emoband_2023}. Most of existing haptic feedback approaches are, however, limited to cutaneous touch}. Gut sensations \change{directly represent inner feelings, allowing designers} to create interactive experiences that communicate inner feelings between individuals, enabling people communicating across distances to share embodied feelings. 

\subsubsection*{\change{Enhancing Empathy with Gastric Sensations: }}

\change{
Our studies demonstrate that gut-based audio-haptic feedback was perceived as realistic, as if originating from participants' own bodies. This quality opens new possibilities for affective communication previously unattainable. For instance, enabling parents to sense how pre-verbal infants feel, or allowing family and friends of autistic individuals to experience their physical sensations. Given that empathy involves neural circuits that activate when observing others' experiences\cite{singer_social_2009}, this technology's ability to transmit visceral sensations may create novel pathways for empathic connection. Future work could investigate how directly experiencing others' internal bodily states influences empathic responses. }

\subsubsection*{Modulating Gastric Interoception for Health and Well-being Applications: } The interest in modulating gastric interoception stems from its critical role in health and well-being, especially eating-related behaviors. The feeling of hunger and fullness influenced the desire to eat \cite{stevenson_psychological_2024}, therefore, artificially influencing the affective associations between food and gastric feelings has significant applications such as shaping eating habits \cite{srinivasan_vibrating_2023, payne_bioelectric_2019}, \change{speeding up recovery in individuals that lost appetite due to medical procedure, or supporting weight management}. \change{Our results suggest that hunger feelings induced by gut-based audio-haptic feedback may change the perceived affective states, as well as their water-drinking behavior.} This implies that gut-based audio-haptic feedback is a promising design element that can be incorporated \change{into} new systems that nudge users. Given that research consistently links impaired gastric interoception with eating disorders \cite{martin_Interoception_2019} -- where individuals struggle to accurately sense hunger or fullness -- our findings open new intervention possibilities. Future work could personalize haptic patterns based on individual gastric sound profiles to dynamically simulate hunger or satiation, helping users better regulate eating behaviors.

\subsubsection*{\revise{Influencing Subjective Food Experiences Through Interoceptive Feedback: }}
\revise{Beyond modulating hunger and fullness, gut-based audio-haptic feedback opens new avenues for investigating how interoceptive signals shape subjective food experiences. Prior work has demonstrated that multisensory cues--such as visual presentation, auditory elements, and textures\cite{spence_extrinsic_2019, johnson_eating_2013}--can significantly influence taste perception, flavor intensity, and food enjoyment. However, the role of interoceptive gastric sensations in shaping these experiences remains largely unexplored}

\revise{Findings from Study 3 suggest that artificially induced gastric sensations could modulate gastric interoceptive behaviors. Therefore, it is possible that these sensations also alter food perception. For instance, hunger sensations might enhance flavor intensity or increase food appeal. This new capability allows future research to further investigate the psychological relationship between gastric interoceptive cues and food perception, creating immersive multisensory dining experiences that synchronize gustatory, olfactory, and interoceptive feedback or developing therapeutic interventions for individuals with altered taste perception due to medical treatments.}

\section{Limitations}
Our study has several limitations. First, \revise{our pool of stimuli is relatively small, with only 3 stimuli per category}. The segmentation of gut sounds is prone to errors, as the first author, despite working with gut sounds intensively, does not have any medical training. Thus, we only \change{labeled} sounds that we were \change{completely sure that they belong} to one of the four categories. \revise{Moreover, we only chose segments from the second half of the WLT, which has lower sound occurrent frequency, which could have impacts on the results.}

Second, the sample size of study 1 is small (N=18), so each stimulus was only experienced by 6 people. This limits the generalizability of our findings, as the experiences created by gut-based haptic feedback are proven to be highly subjective. 

Finally, study 3 used the Water Load Test-II to quantify gastric interoceptive behaviors. However, we acknowledge that drinking volumes might be influenced by various factors. Although we have employed measures to minimize the effect, such as asking participants to come at the same time every day, as well as using a within-subject study design to control individual differences, there could still be confounding factors that we did not account for. Moreover, we sacrificed generalizability to maximize the chance to detect significance by choosing two contrasting sub-segments of the stimuli categories. \change{Furthermore, while we successfully demonstrated} the possibility of non-invasively manipulating gastric interoception, \change{the results were counterintuitive}: the least hunger-inducing segment increased the amount of ingested water. Although we provided hypotheses to explain the unexpected outcome, the phenomenon should be further investigated in the future. 

\section{Conclusion}
Our work provides the first empirical evidence that gut-based audio-haptic feedback can non-invasively modulate gastric interoception. We showed that such feedback not only elicits visceral gastric feelings but also alters perceived hunger and satiety levels, with measurable effects on drinking behavior. Our findings open up new possibilities for HCI by introducing gastric interoception as an interaction modality, offering potential applications in health, well-being, and behavioral nudging. 

\begin{acks}
  We would like to thank Himeth Thawmika and Le Nguyen for their outstanding assistance in running back-to-back user studies from 9am to 7pm. We thank Vanessa Tang for her modeling support, and the members of the Augmented Human Lab who volunteered in the various pilot studies that informed the work presented in this paper. We are grateful to Yasith Sam for discussions during earlier iterations of this project, and to KP Yeo for his selfless and generous technical support. We also thank the participants for meeting the strict study prerequisites and for providing valuable feedback and insights. Finally, we thank Hung Viet Nguyen for being a constant source of encouragement and for providing Mia Huong Nguyen with much-needed confidence boosters.
This research is partially supported by NUS Smart System Institute foresight grant for Suranga Nanayakkara. Mia Huong Nguyen is supported by the NUS-SINGA Award. 
  
\end{acks}

\bibliographystyle{ACM-Reference-Format}
\bibliography{myzot}
\appendix
\section{Appendix}

\begin{figure*}[h!]
    \centering
    \includegraphics[width=1\linewidth]{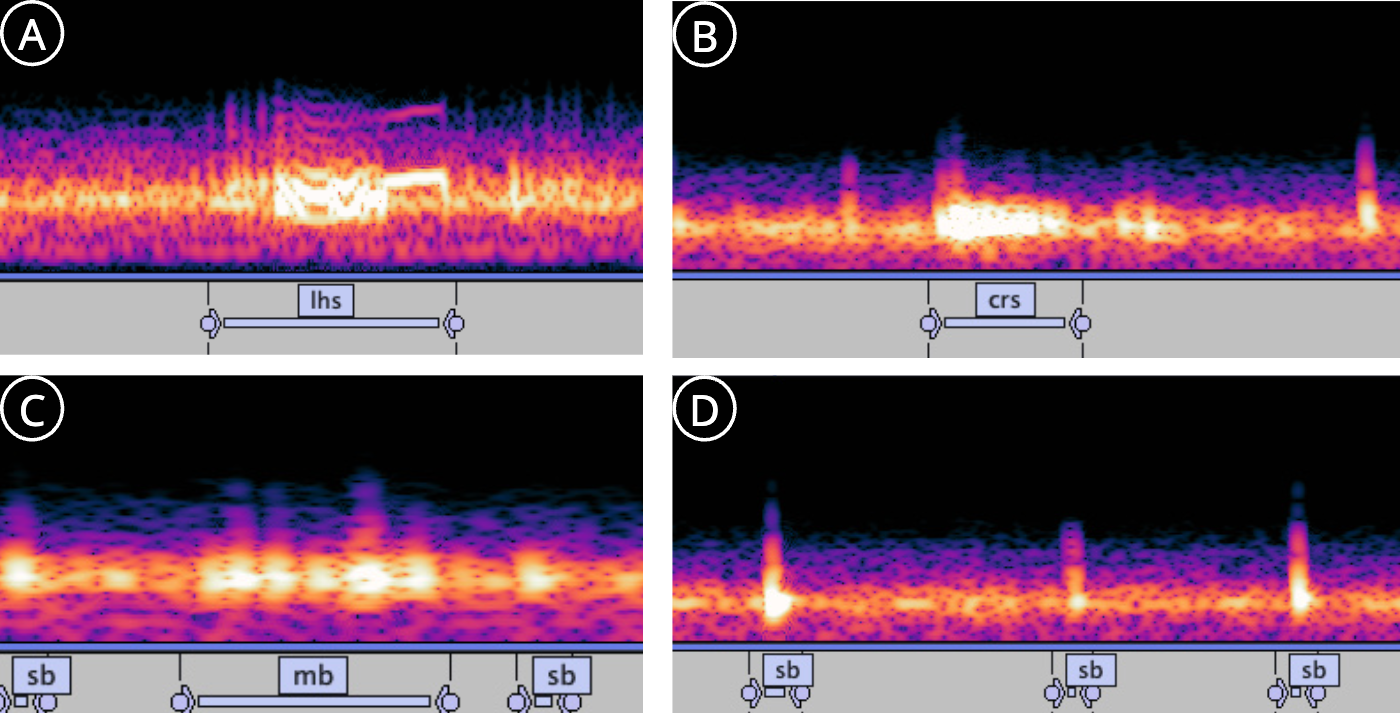}
    \caption{Examples of bowel sounds in 4 categories: \Circled{A}: Long Harmonic Sounds(lhs), \Circled{B}: Continuous Random Sounds (crs), \Circled{C}: Multiple Bursts (mb) and Single Burst (sb), \Circled{D}: Single Burst (sb)}
    \label{fig:example}
    \Description{
    The figure consists of four spectrograms arranged in a 2×2 grid, labeled (A) through (D). In each panel, the horizontal axis represents time and the vertical axis represents frequency, with color indicating signal intensity from low (dark purple/blue) to high (bright yellow/white). 
    (A) LHS shows a relatively continuous horizontal band of energy with moderate fluctuations and a brief region of increased intensity near the center. 
    (B) CRS exhibits a steady background band punctuated by several narrow, high-intensity vertical spikes, indicating short, transient events. 
    (C) MB, flanked by SB segments, displays repeated bursts of high-intensity energy across time, forming a rhythmic pattern with multiple peaks. 
    (D) SB shows sparse, isolated high-intensity vertical spikes separated by longer low-intensity intervals. 
    Together, the panels illustrate distinct temporal and intensity characteristics of the four gut-based audio stimuli.
    }

\end{figure*}
%
%

\end{document}